
\input fontch.tex

%
%
%
\def\unredoffs{} \def\redoffs{\voffset=-.31truein\hoffset=-.48truein}
\def\speclscape{}
%
%
%
%
%
\newbox\leftpage \newdimen\fullhsize \newdimen\hstitle \newdimen\hsbody
\tolerance=1000\hfuzz=2pt
\catcode`\@=11 
\ifx\hyperdef\UNd@FiNeD\def\hyperdef#1#2#3#4{#4}\def\hyperref#1#2#3#4{#4}\fi
\def\bigans{b }
\def\answ{b }
%
\ifx\answ\bigans\message{(This will come out unreduced.}
\magnification=1200\unredoffs\baselineskip=16pt plus 2pt minus 1pt
\hsbody=\hsize \hstitle=\hsize 
\else\message{(This will be reduced.} \let\l@r=L
\magnification=1000\baselineskip=16pt plus 2pt minus 1pt \vsize=7truein
\redoffs \hstitle=8truein\hsbody=4.75truein\fullhsize=10truein\hsize=\hsbody
\output={\ifnum\pageno=0 
  \shipout\vbox{\speclscape{\hsize\fullhsize\makeheadline}
    \hbox to \fullhsize{\hfill\pagebody\hfill}}\advancepageno
  \else
  \almostshipout{\leftline{\vbox{\pagebody\makefootline}}}\advancepageno
  \fi}
\def\almostshipout#1{\if L\l@r \count1=1 \message{[\the\count0.\the\count1]}
      \global\setbox\leftpage=#1 \global\let\l@r=R
 \else \count1=2
  \shipout\vbox{\speclscape{\hsize\fullhsize\makeheadline}
      \hbox to\fullhsize{\box\leftpage\hfil#1}}  \global\let\l@r=L\fi}
\fi
%
\newcount\yearltd\yearltd=\year\advance\yearltd by -2000

\def\Title#1#2{\nopagenumbers\abstractfont\hsize=\hstitle\rightline{#1}%
\vskip 1in\centerline{\titlefont #2}\abstractfont\vskip .5in\pageno=0}
\def\Date#1{\vfill\leftline{#1}\tenpoint\supereject\global\hsize=\hsbody%
\footline={\hss\tenrm\hyperdef\hypernoname{page}\folio\folio\hss}}%
%

\def\draftmode{\message{ DRAFTMODE }\def\draftdate{{\rm preliminary draft:
\number\month/\number\day/\number\yearltd\ \ \hourmin}}%
\headline={\hfil\draftdate}\writelabels\baselineskip=20pt plus 2pt minus 2pt
 {\count255=\time\divide\count255 by 60 \xdef\hourmin{\number\count255}
  \multiply\count255 by-60\advance\count255 by\time
  \xdef\hourmin{\hourmin:\ifnum\count255<10 0\fi\the\count255}}}
\def\nolabels{\def\wrlabeL##1{}\def\eqlabeL##1{}\def\reflabeL##1{}}
\def\writelabels{\def\wrlabeL##1{\leavevmode\vadjust{\rlap{\smash%
{\line{{\escapechar=` \hfill\rlap{\sevenrm\hskip.03in\string##1}}}}}}}%
\def\eqlabeL##1{{\escapechar-1\rlap{\sevenrm\hskip.05in\string##1}}}%
\def\reflabeL##1{\noexpand\llap{\noexpand\sevenrm\string\string\string##1}}}
\nolabels
%
\global\newcount\secno \global\secno=0
\global\newcount\meqno \global\meqno=1
\def\s@csym{}
\def\newsec#1{\global\advance\secno by1%
{\toks0{#1}\message{(\the\secno. \the\toks0)}}%
\global\subsecno=0\eqnres@t\let\s@csym\secsym\xdef\secn@m{\the\secno}\noindent
{\bf\hyperdef\hypernoname{section}{\the\secno}{\the\secno.} #1}%
\writetoca{{\string\hyperref{}{section}{\the\secno}{\the\secno.}} {#1}}%
\par\nobreak\medskip\nobreak}
\def\eqnres@t{\xdef\secsym{\the\secno.}\global\meqno=1\bigbreak\bigskip}
\def\sequentialequations{\def\eqnres@t{\bigbreak}}\xdef\secsym{}
\global\newcount\subsecno \global\subsecno=0
\def\subsec#1{\global\advance\subsecno by1%
{\toks0{#1}\message{(\s@csym\the\subsecno. \the\toks0)}}%
\ifnum\lastpenalty>9000\else\bigbreak\fi
\noindent{\it\hyperdef\hypernoname{subsection}{\secn@m.\the\subsecno}%
{\secn@m.\the\subsecno.} #1}\writetoca{\string\quad
{\string\hyperref{}{subsection}{\secn@m.\the\subsecno}{\secn@m.\the\subsecno.}}
{#1}}\par\nobreak\medskip\nobreak}
\def\appendix#1#2{\global\meqno=1\global\subsecno=0\xdef\secsym{\hbox{#1.}}%
\bigbreak\bigskip\noindent{\bf Appendix \hyperdef\hypernoname{appendix}{#1}%
{#1.} #2}{\toks0{(#1. #2)}\message{\the\toks0}}%
\xdef\s@csym{#1.}\xdef\secn@m{#1}%
\writetoca{\string\hyperref{}{appendix}{#1}{Appendix {#1.}} {#2}}%
\par\nobreak\medskip\nobreak}
%
%
\def\checkm@de#1#2{\ifmmode{\def\f@rst##1{##1}\hyperdef\hypernoname{equation}%
{#1}{#2}}\else\hyperref{}{equation}{#1}{#2}\fi}
\def\eqnn#1{\DefWarn#1\xdef #1{(\noexpand\relax\noexpand\checkm@de%
{\s@csym\the\meqno}{\secsym\the\meqno})}%
\wrlabeL#1\writedef{#1\leftbracket#1}\global\advance\meqno by1}
\def\f@rst#1{\c@t#1a\em@ark}\def\c@t#1#2\em@ark{#1}
\def\eqna#1{\DefWarn#1\wrlabeL{#1$\{\}$}%
\xdef #1##1{(\noexpand\relax\noexpand\checkm@de%
{\s@csym\the\meqno\noexpand\f@rst{##1}}{\hbox{$\secsym\the\meqno##1$}})}
\writedef{#1\numbersign1\leftbracket#1{\numbersign1}}\global\advance\meqno by1}
\def\eqn#1#2{\DefWarn#1%
\xdef #1{(\noexpand\hyperref{}{equation}{\s@csym\the\meqno}%
{\secsym\the\meqno})}$$#2\eqno(\hyperdef\hypernoname{equation}%
{\s@csym\the\meqno}{\secsym\the\meqno})\eqlabeL#1$$%
\writedef{#1\leftbracket#1}\global\advance\meqno by1}
\def\xeqn{\expandafter\xe@n}\def\xe@n(#1){#1}
\def\xeqna#1{\expandafter\xe@n#1}
\def\eqns#1{(\e@ns #1{\hbox{}})}
\def\e@ns#1{\ifx\UNd@FiNeD#1\message{eqnlabel \string#1 is undefined.}%
\xdef#1{(?.?)}\fi{\let\hyperref=\relax\xdef\next{#1}}%
\ifx\next\em@rk\def\next{}\else%
\ifx\next#1\xeqn#1\else\def\n@xt{#1}\ifx\n@xt\next#1\else\xeqna#1\fi
\fi\let\next=\e@ns\fi\next}

\def\DefWarn#1{\ifx\UNd@FiNeD#1\else
\immediate\write16{*** WARNING: the label \string#1 is already defined ***}\fi}
%
\newskip\footskip\footskip14pt plus 1pt minus 1pt 
\def\footnotefont{\ninepoint}\def\f@t#1{\footnotefont #1\@foot}
\def\f@@t{\baselineskip\footskip\bgroup\footnotefont\aftergroup\@foot\let\next}
\setbox\strutbox=\hbox{\vrule height9.5pt depth4.5pt width0pt}
\global\newcount\ftno \global\ftno=0
\def\foot{\global\advance\ftno by1\def\foot@rg{\hyperref{}{footnote}%
{\the\ftno}{\the\ftno}\xdef\foot@rg{\noexpand\hyperdef\noexpand\hypernoname%
{footnote}{\the\ftno}{\the\ftno}}}\footnote{$^{\foot@rg}$}}
%
\newwrite\ftfile
\def\footend{\def\foot{\global\advance\ftno by1\chardef\wfile=\ftfile
\hyperref{}{footnote}{\the\ftno}{$^{\the\ftno}$}%
\ifnum\ftno=1\immediate\openout\ftfile=\jobname.fts\fi%
\immediate\write\ftfile{\noexpand\smallskip%
\noexpand\item{\noexpand\hyperdef\noexpand\hypernoname{footnote}
{\the\ftno}{f\the\ftno}:\ }\pctsign}\findarg}%
\def\footatend{\vfill\eject\immediate\closeout\ftfile{\parindent=20pt
\centerline{\bf Footnotes}\nobreak\bigskip\input \jobname.fts }}}
\def\footatend{}
%
%
\global\newcount\refno \global\refno=1
\newwrite\rfile
\def\ref{[\hyperref{}{reference}{\the\refno}{\the\refno}]\nref}
\def\nref#1{\DefWarn#1%
\xdef#1{[\noexpand\hyperref{}{reference}{\the\refno}{\the\refno}]}%
\writedef{#1\leftbracket#1}%
\ifnum\refno=1\immediate\openout\rfile=\jobname.refs\fi
\chardef\wfile=\rfile\immediate\write\rfile{\noexpand\item{[\noexpand\hyperdef%
\noexpand\hypernoname{reference}{\the\refno}{\the\refno}]\ }%
\reflabeL{#1\hskip.31in}\pctsign}\global\advance\refno by1\findarg}
\def\findarg#1#{\begingroup\obeylines\newlinechar=`\^^M\pass@rg}
{\obeylines\gdef\pass@rg#1{\writ@line\relax #1^^M\hbox{}^^M}%
\gdef\writ@line#1^^M{\expandafter\toks0\expandafter{\striprel@x #1}%
\edef\next{\the\toks0}\ifx\next\em@rk\let\next=\endgroup\else\ifx\next\empty%
\else\immediate\write\wfile{\the\toks0}\fi\let\next=\writ@line\fi\next\relax}}
\def\striprel@x#1{} \def\em@rk{\hbox{}}
\def\lref{\begingroup\obeylines\lr@f}
\def\lr@f#1#2{\DefWarn#1\gdef#1{\let#1=\UNd@FiNeD\ref#1{#2}}\endgroup\unskip}

\def\addref#1{\immediate\write\rfile{\noexpand\item{}#1}} 
\def\listrefs{\footatend\vfill\supereject\immediate\closeout\rfile\writestoppt
\baselineskip=\footskip\centerline{{\bf References}}\bigskip{\parindent=20pt%
\frenchspacing\escapechar=` \input \jobname.refs\vfill\eject}\nonfrenchspacing}
\def\startrefs#1{\immediate\openout\rfile=\jobname.refs\refno=#1}
\def\xref{\expandafter\xr@f}\def\xr@f[#1]{#1}
\def\refs#1{\count255=1[\r@fs #1{\hbox{}}]}
\def\r@fs#1{\ifx\UNd@FiNeD#1\message{reflabel \string#1 is undefined.}%
\nref#1{need to supply reference \string#1.}\fi%
\vphantom{\hphantom{#1}}{\let\hyperref=\relax\xdef\next{#1}}%
\ifx\next\em@rk\def\next{}%
\else\ifx\next#1\ifodd\count255\relax\xref#1\count255=0\fi%
\else#1\count255=1\fi\let\next=\r@fs\fi\next}
%

%
\newwrite\ffile\global\newcount\figno \global\figno=1
\def\fig{fig.~\hyperref{}{figure}{\the\figno}{\the\figno}\nfig}
\def\nfig#1{\DefWarn#1%
\xdef#1{fig.~\noexpand\hyperref{}{figure}{\the\figno}{\the\figno}}%
\writedef{#1\leftbracket fig.\noexpand~\xfig#1}%
\ifnum\figno=1\immediate\openout\ffile=\jobname.figs\fi\chardef\wfile=\ffile%
{\let\hyperref=\relax
\immediate\write\ffile{\noexpand\medskip\noexpand\item{Fig.\ %
\noexpand\hyperdef\noexpand\hypernoname{figure}{\the\figno}{\the\figno}. }
\reflabeL{#1\hskip.55in}\pctsign}}\global\advance\figno by1\findarg}
\def\listfigs{\vfill\eject\immediate\closeout\ffile{\parindent40pt
\baselineskip14pt\centerline{{\bf Figure Captions}}\nobreak\medskip
\escapechar=` \input \jobname.figs\vfill\eject}}
\def\xfig{\expandafter\xf@g}\def\xf@g fig.\penalty\@M\ {}
\def\figs#1{figs.~\f@gs #1{\hbox{}}}
\def\f@gs#1{{\let\hyperref=\relax\xdef\next{#1}}\ifx\next\em@rk\def\next{}\else
\ifx\next#1\xfig #1\else#1\fi\let\next=\f@gs\fi\next}
\def\figin{\epsfcheck\figin}\def\figins{\epsfcheck\figins}
\def\epsfcheck{\ifx\epsfbox\UNd@FiNeD
\message{(NO epsf.tex, FIGURES WILL BE IGNORED)}
\gdef\figin##1{\vskip2in}\gdef\figins##1{\hskip.5in}
\else\message{(FIGURES WILL BE INCLUDED)}%
\gdef\figin##1{##1}\gdef\figins##1{##1}\fi}
\def\DefWarn#1{}
\def\figinsert{\goodbreak\midinsert}
\def\ifig#1#2#3{\DefWarn#1\xdef#1{fig.~\noexpand\hyperref{}{figure}%
{\the\figno}{\the\figno}}\writedef{#1\leftbracket fig.\noexpand~\xfig#1}%
\figinsert\figin{\centerline{#3}}\medskip\centerline{\vbox{\baselineskip12pt
\advance\hsize by -1truein\noindent\wrlabeL{#1=#1}\footnotefont%
{\bf Fig.~\hyperdef\hypernoname{figure}{\the\figno}{\the\figno}:} #2}}
\bigskip\endinsert\global\advance\figno by1}
\newwrite\lfile
{\escapechar-1\xdef\pctsign{\string\%}\xdef\leftbracket{\string\{}
\xdef\rightbracket{\string\}}\xdef\numbersign{\string\#}}
\def\writedefs{\immediate\openout\lfile=\jobname.defs \def\writedef##1{%
{\let\hyperref=\relax\let\hyperdef=\relax\let\hypernoname=\relax
 \immediate\write\lfile{\string\def\string##1\rightbracket}}}}%
\def\writestop{\def\writestoppt{\immediate\write\lfile{\string\pageno
 \the\pageno\string\startrefs\leftbracket\the\refno\rightbracket
 \string\def\string\secsym\leftbracket\secsym\rightbracket
 \string\secno\the\secno\string\meqno\the\meqno}\immediate\closeout\lfile}}
\def\writestoppt{}\def\writedef#1{}
\def\seclab#1{\DefWarn#1%
\xdef #1{\noexpand\hyperref{}{section}{\the\secno}{\the\secno}}%
\writedef{#1\leftbracket#1}\wrlabeL{#1=#1}}
\def\subseclab#1{\DefWarn#1%
\xdef #1{\noexpand\hyperref{}{subsection}{\secn@m.\the\subsecno}%
{\secn@m.\the\subsecno}}\writedef{#1\leftbracket#1}\wrlabeL{#1=#1}}
\def\applab#1{\DefWarn#1%
\xdef #1{\noexpand\hyperref{}{appendix}{\secn@m}{\secn@m}}%
\writedef{#1\leftbracket#1}\wrlabeL{#1=#1}}
\newwrite\tfile \def\writetoca#1{}
\def\leaderfill{\leaders\hbox to 1em{\hss.\hss}\hfill}
\def\writetoc{\immediate\openout\tfile=\jobname.toc
   \def\writetoca##1{{\edef\next{\write\tfile{\noindent ##1
   \string\leaderfill {\string\hyperref{}{page}{\noexpand\number\pageno}%
                       {\noexpand\number\pageno}} \par}}\next}}}
\newread\ch@ckfile
\def\listtoc{\immediate\closeout\tfile\immediate\openin\ch@ckfile=\jobname.toc
\ifeof\ch@ckfile\message{no file \jobname.toc, no table of contents this pass}%
\else\closein\ch@ckfile\centerline{\bf Contents}\nobreak\medskip%
{\baselineskip=12pt\footnotefont\parskip=0pt\catcode`\@=11\input\jobname.toc
\catcode`\@=12\bigbreak\bigskip}\fi}
\catcode`\@=12 
%
\edef\tfontsize{\ifx\answ\bigans scaled\magstep3\else scaled\magstep4\fi}
\font\titlerm=cmr10 \tfontsize \font\titlerms=cmr7 \tfontsize
\font\titlermss=cmr5 \tfontsize \font\titlei=cmmi10 \tfontsize
\font\titleis=cmmi7 \tfontsize \font\titleiss=cmmi5 \tfontsize
\font\titlesy=cmsy10 \tfontsize \font\titlesys=cmsy7 \tfontsize
\font\titlesyss=cmsy5 \tfontsize \font\titleit=cmti10 \tfontsize
\skewchar\titlei='177 \skewchar\titleis='177 \skewchar\titleiss='177
\skewchar\titlesy='60 \skewchar\titlesys='60 \skewchar\titlesyss='60
\def\titlefont{\def\rm{\fam0\titlerm}
\textfont0=\titlerm \scriptfont0=\titlerms \scriptscriptfont0=\titlermss
\textfont1=\titlei \scriptfont1=\titleis \scriptscriptfont1=\titleiss
\textfont2=\titlesy \scriptfont2=\titlesys \scriptscriptfont2=\titlesyss
\textfont\itfam=\titleit \def\it{\fam\itfam\titleit}\rm}
 \ifx\answ\bigans\else scaled\magstep1\fi
\ifx\answ\bigans\def\abstractfont{\tenpoint}\else
\font\absit=cmti10 scaled \magstep1
\font\abssl=cmsl10 scaled \magstep1
\font\absrm=cmr10 scaled\magstep1 \font\absrms=cmr7 scaled\magstep1
\font\absrmss=cmr5 scaled\magstep1 \font\absi=cmmi10 scaled\magstep1
\font\absis=cmmi7 scaled\magstep1 \font\absiss=cmmi5 scaled\magstep1
\font\abssy=cmsy10 scaled\magstep1 \font\abssys=cmsy7 scaled\magstep1
\font\abssyss=cmsy5 scaled\magstep1 \font\absbf=cmbx10 scaled\magstep1
\skewchar\absi='177 \skewchar\absis='177 \skewchar\absiss='177
\skewchar\abssy='60 \skewchar\abssys='60 \skewchar\abssyss='60
\def\abstractfont{\def\rm{\fam0\absrm}
\textfont0=\absrm \scriptfont0=\absrms \scriptscriptfont0=\absrmss
\textfont1=\absi \scriptfont1=\absis \scriptscriptfont1=\absiss
\textfont2=\abssy \scriptfont2=\abssys \scriptscriptfont2=\abssyss
\textfont\itfam=\absit \def\it{\fam\itfam\absit}\def\footnotefont{\tenpoint}%
\textfont\slfam=\abssl \def\sl{\fam\slfam\abssl}%
\textfont\bffam=\absbf \def\bf{\fam\bffam\absbf}\rm}\fi
\def\tenpoint{\def\rm{\fam0\tenrm}
\textfont0=\tenrm \scriptfont0=\sevenrm \scriptscriptfont0=\fiverm
\textfont1=\teni  \scriptfont1=\seveni  \scriptscriptfont1=\fivei
\textfont2=\tensy \scriptfont2=\sevensy \scriptscriptfont2=\fivesy
\textfont\itfam=\tenit \def\it{\fam\itfam\tenit}\def\footnotefont{\ninepoint}%
\textfont\bffam=\tenbf \def\bf{\fam\bffam\tenbf}\def\sl{\fam\slfam\tensl}\rm}
\font\ninerm=cmr9 \font\sixrm=cmr6 \font\ninei=cmmi9 \font\sixi=cmmi6
\font\ninesy=cmsy9 \font\sixsy=cmsy6 \font\ninebf=cmbx9
\font\nineit=cmti9 \font\ninesl=cmsl9 \skewchar\ninei='177
\skewchar\sixi='177 \skewchar\ninesy='60 \skewchar\sixsy='60
\def\ninepoint{\def\rm{\fam0\ninerm}
\textfont0=\ninerm \scriptfont0=\sixrm \scriptscriptfont0=\fiverm
\textfont1=\ninei \scriptfont1=\sixi \scriptscriptfont1=\fivei
\textfont2=\ninesy \scriptfont2=\sixsy \scriptscriptfont2=\fivesy
\textfont\itfam=\ninei \def\it{\fam\itfam\nineit}\def\sl{\fam\slfam\ninesl}%
\textfont\bffam=\ninebf \def\bf{\fam\bffam\ninebf}\rm}
%
%

\hyphenation{anom-aly anom-alies coun-ter-term coun-ter-terms}
\def\inv{^{\raise.15ex\hbox{${\scriptscriptstyle -}$}\kern-.05em 1}}

\def\Dsl{\,\raise.15ex\hbox{/}\mkern-13.5mu D} 
\def\dsl{\raise.15ex\hbox{/}\kern-.57em\partial}

 \def\Tr{{\rm Tr}}
\def\lspace{\ifx\answ\bigans{}\else\qquad\fi}
\def\lbspace{\ifx\answ\bigans{}\else\hskip-.2in\fi} 
\def\boxeqn#1{\vcenter{\vbox{\hrule\hbox{\vrule\kern3pt\vbox{\kern3pt
	\hbox{${\displaystyle #1}$}\kern3pt}\kern3pt\vrule}\hrule}}}
\def\mbox#1#2{\vcenter{\hrule \hbox{\vrule height#2in
		\kern#1in \vrule} \hrule}}  
%

\def\darr#1{\raise1.5ex\hbox{$\leftrightarrow$}\mkern-16.5mu #1}

\def\roughly#1{\raise.3ex\hbox{$#1$\kern-.75em\lower1ex\hbox{$\sim$}}}

\def\bb{
\font\tenmsb=msbm10
\font\sevenmsb=msbm7
\font\fivemsb=msbm5
\textfont1=\tenmsb
\scriptfont1=\sevenmsb
\scriptscriptfont1=\fivemsb
}

\input amssym

\input epsf

\def\IZ{\relax\ifmmode\mathchoice
{\hbox{\cmss Z\kern-.4em Z}}{\hbox{\cmss Z\kern-.4em Z}} {\lower.9pt\hbox{\cmsss Z\kern-.4em Z}}
{\lower1.2pt\hbox{\cmsss Z\kern-.4em Z}}\else{\cmss Z\kern-.4em Z}\fi}

\newif\ifdraft\draftfalse
\newif\ifinter\interfalse
\ifdraft\draftmode\else\interfalse\fi
\def\journal#1&#2(#3){\unskip, \sl #1\ \bf #2 \rm(19#3) }
\def\andjournal#1&#2(#3){\sl #1~\bf #2 \rm (19#3) }

\def\frac#1#2{{#1\over#2}}

\def\inbar{\,\vrule height1.5ex width.4pt depth0pt}
\def\IC{\relax\hbox{$\inbar\kern-.3em{\rm C}$}}
\def\IR{\relax{\rm I\kern-.18em R}}
\def\IP{\relax{\rm I\kern-.18em P}}
\def\Z{{\bf Z}}

%
%


%
\catcode`\@=11
\def\slash#1{\mathord{\mathpalette\c@ncel{#1}}}
\overfullrule=0pt

\def\II{{\cal I}}

\def\underrel#1\over#2{\mathrel{\mathop{\kern\z@#1}\limits_{#2}}}

\catcode`\@=12


%

\def\exp{{\rm exp}}


\def\[{[}
\def\]{]}

\def\comment#1{ }

%
\def\draftnote#1{\ifdraft{\baselineskip2ex
                 \vbox{\kern1em\hrule\hbox{\vrule\kern1em\vbox{\kern1ex
                 \noindent \underbar{NOTE}: #1
             \vskip1ex}\kern1em\vrule}\hrule}}\fi}
\def\internote#1{\ifinter{\baselineskip2ex
                 \vbox{\kern1em\hrule\hbox{\vrule\kern1em\vbox{\kern1ex
                 \noindent \underbar{Internal Note}: #1
             \vskip1ex}\kern1em\vrule}\hrule}}\fi}

%
%



%
%
%
%

%

\def\inv{^{-1}}


\def\Tr{{\rm Tr}}

\def\1{{\ds 1}}
\def\R{\hbox{$\bb R$}}
\def\C{\hbox{$\bb C$}}

\def\Z{\hbox{$\bb Z$}}

\def\P{\hbox{$\bb P$}}

\def\S{\hbox{$\bb S$}}

\newfam\frakfam
\font\teneufm=eufm10
\font\seveneufm=eufm7
\font\fiveeufm=eufm5
\textfont\frakfam=\teneufm
\scriptfont\frakfam=\seveneufm
\scriptscriptfont\frakfam=\fiveeufm

\lref\NiarchosAH{
  V.~Niarchos,
  ``Seiberg dualities and the 3d/4d connection,''
JHEP {\bf 1207}, 075 (2012).
[arXiv:1205.2086 [hep-th]].
}

\lref\AharonyGP{
  O.~Aharony,
  ``IR duality in d = 3 N=2 supersymmetric USp(2N(c)) and U(N(c)) gauge theories,''
Phys.\ Lett.\ B {\bf 404}, 71 (1997).
[hep-th/9703215].
}

\lref\AffleckAS{
  I.~Affleck, J.~A.~Harvey and E.~Witten,
  ``Instantons and (Super)Symmetry Breaking in (2+1)-Dimensions,''
Nucl.\ Phys.\ B {\bf 206}, 413 (1982)..
}

\lref\IntriligatorID{
  K.~A.~Intriligator and N.~Seiberg,
  ``Duality, monopoles, dyons, confinement and oblique confinement in supersymmetric SO(N(c)) gauge theories,''
Nucl.\ Phys.\ B {\bf 444}, 125 (1995).
[hep-th/9503179].
}

\lref\PasquettiFJ{
  S.~Pasquetti,
  ``Factorisation of N = 2 Theories on the Squashed 3-Sphere,''
JHEP {\bf 1204}, 120 (2012).
[arXiv:1111.6905 [hep-th]].
}

\lref\BeemMB{
  C.~Beem, T.~Dimofte and S.~Pasquetti,
  ``Holomorphic Blocks in Three Dimensions,''
[arXiv:1211.1986 [hep-th]].
}

\lref\SeibergPQ{
  N.~Seiberg,
  ``Electric - magnetic duality in supersymmetric nonAbelian gauge theories,''
Nucl.\ Phys.\ B {\bf 435}, 129 (1995).
[hep-th/9411149].
}

\lref\AharonyBX{
  O.~Aharony, A.~Hanany, K.~A.~Intriligator, N.~Seiberg and M.~J.~Strassler,
  ``Aspects of N=2 supersymmetric gauge theories in three-dimensions,''
Nucl.\ Phys.\ B {\bf 499}, 67 (1997).
[hep-th/9703110].
}

\lref\IntriligatorNE{
  K.~A.~Intriligator and P.~Pouliot,
  ``Exact superpotentials, quantum vacua and duality in supersymmetric SP(N(c)) gauge theories,''
Phys.\ Lett.\ B {\bf 353}, 471 (1995).
[hep-th/9505006].
}

\lref\KarchUX{
  A.~Karch,
  ``Seiberg duality in three-dimensions,''
Phys.\ Lett.\ B {\bf 405}, 79 (1997).
[hep-th/9703172].
}

\lref\SafdiRE{
  B.~R.~Safdi, I.~R.~Klebanov and J.~Lee,
  ``A Crack in the Conformal Window,''
[arXiv:1212.4502 [hep-th]].
}

\lref\SchweigertTG{
  C.~Schweigert,
  ``On moduli spaces of flat connections with nonsimply connected structure group,''
Nucl.\ Phys.\ B {\bf 492}, 743 (1997).
[hep-th/9611092].
}

\lref\GiveonZN{
  A.~Giveon and D.~Kutasov,
  ``Seiberg Duality in Chern-Simons Theory,''
Nucl.\ Phys.\ B {\bf 812}, 1 (2009).
[arXiv:0808.0360 [hep-th]].
}

\lref\GaiottoBE{
  D.~Gaiotto, G.~W.~Moore and A.~Neitzke,
  ``Framed BPS States,''
[arXiv:1006.0146 [hep-th]].
}

\lref\AldayRS{
  L.~F.~Alday, M.~Bullimore and M.~Fluder,
  ``On S-duality of the Superconformal Index on Lens Spaces and 2d TQFT,''
JHEP {\bf 1305}, 122 (2013).
[arXiv:1301.7486 [hep-th]].
}

\lref\RazamatJXA{
  S.~S.~Razamat and M.~Yamazaki,
  ``S-duality and the N=2 Lens Space Index,''
[arXiv:1306.1543 [hep-th]].
}

\lref\NiarchosAH{
  V.~Niarchos,
  ``Seiberg dualities and the 3d/4d connection,''
JHEP {\bf 1207}, 075 (2012).
[arXiv:1205.2086 [hep-th]].
}

\lref\almost{
  A.~Borel, R.~Friedman, J.~W.~Morgan,
  ``Almost commuting elements in compact Lie groups,''
arXiv:math/9907007.
}

\lref\KapustinJM{
  A.~Kapustin and B.~Willett,
  ``Generalized Superconformal Index for Three Dimensional Field Theories,''
[arXiv:1106.2484 [hep-th]].
}

\lref\AharonyGP{
  O.~Aharony,
  ``IR duality in d = 3 N=2 supersymmetric USp(2N(c)) and U(N(c)) gauge theories,''
Phys.\ Lett.\ B {\bf 404}, 71 (1997).
[hep-th/9703215].
}

\lref\FestucciaWS{
  G.~Festuccia and N.~Seiberg,
  ``Rigid Supersymmetric Theories in Curved Superspace,''
JHEP {\bf 1106}, 114 (2011).
[arXiv:1105.0689 [hep-th]].
}

\lref\RomelsbergerEG{
  C.~Romelsberger,
  ``Counting chiral primaries in N = 1, d=4 superconformal field theories,''
Nucl.\ Phys.\ B {\bf 747}, 329 (2006).
[hep-th/0510060].
}

\lref\KapustinKZ{
  A.~Kapustin, B.~Willett and I.~Yaakov,
  ``Exact Results for Wilson Loops in Superconformal Chern-Simons Theories with Matter,''
JHEP {\bf 1003}, 089 (2010).
[arXiv:0909.4559 [hep-th]].
}

\lref\DolanQI{
  F.~A.~Dolan and H.~Osborn,
  ``Applications of the Superconformal Index for Protected Operators and q-Hypergeometric Identities to N=1 Dual Theories,''
Nucl.\ Phys.\ B {\bf 818}, 137 (2009).
[arXiv:0801.4947 [hep-th]].
}

\lref\GaddeIA{
  A.~Gadde and W.~Yan,
  ``Reducing the 4d Index to the $S^3$ Partition Function,''
JHEP {\bf 1212}, 003 (2012).
[arXiv:1104.2592 [hep-th]].
}

\lref\DolanRP{
  F.~A.~H.~Dolan, V.~P.~Spiridonov and G.~S.~Vartanov,
  ``From 4d superconformal indices to 3d partition functions,''
Phys.\ Lett.\ B {\bf 704}, 234 (2011).
[arXiv:1104.1787 [hep-th]].
}

\lref\ImamuraUW{
  Y.~Imamura,
 ``Relation between the 4d superconformal index and the $S^3$ partition function,''
JHEP {\bf 1109}, 133 (2011).
[arXiv:1104.4482 [hep-th]].
}

\lref\HamaEA{
  N.~Hama, K.~Hosomichi and S.~Lee,
  ``SUSY Gauge Theories on Squashed Three-Spheres,''
JHEP {\bf 1105}, 014 (2011).
[arXiv:1102.4716 [hep-th]].
}

\lref\GaddeEN{
  A.~Gadde, L.~Rastelli, S.~S.~Razamat and W.~Yan,
  ``On the Superconformal Index of N=1 IR Fixed Points: A Holographic Check,''
JHEP {\bf 1103}, 041 (2011).
[arXiv:1011.5278 [hep-th]].
}

\lref\EagerHX{
  R.~Eager, J.~Schmude and Y.~Tachikawa,
  ``Superconformal Indices, Sasaki-Einstein Manifolds, and Cyclic Homologies,''
[arXiv:1207.0573 [hep-th]].
}

\lref\AffleckAS{
  I.~Affleck, J.~A.~Harvey and E.~Witten,
  ``Instantons and (Super)Symmetry Breaking in (2+1)-Dimensions,''
Nucl.\ Phys.\ B {\bf 206}, 413 (1982)..
}

\lref\SeibergPQ{
  N.~Seiberg,
  ``Electric - magnetic duality in supersymmetric nonAbelian gauge theories,''
Nucl.\ Phys.\ B {\bf 435}, 129 (1995).
[hep-th/9411149].
}

\lref\debult{
  F.~van~de~Bult,
  ``Hyperbolic Hypergeometric Functions,''
University of Amsterdam Ph.D. thesis
}

\lref\Shamirthesis{
  I.~Shamir,
  ``Aspects of three dimensional Seiberg duality,''
  M. Sc. thesis submitted to the Weizmann Institute of Science, April 2010.
  }

\lref\slthreeZ{
  J.~Felder, A.~Varchenko,
  ``The elliptic gamma function and $SL(3,Z) \times Z^3$,'' $\;\;$
[arXiv:math/0001184].
}

\lref\BeniniNC{
  F.~Benini, T.~Nishioka and M.~Yamazaki,
  ``4d Index to 3d Index and 2d TQFT,''
Phys.\ Rev.\ D {\bf 86}, 065015 (2012).
[arXiv:1109.0283 [hep-th]].
}

\lref\GaiottoWE{
  D.~Gaiotto,
  ``N=2 dualities,''
  JHEP {\bf 1208}, 034 (2012).
  [arXiv:0904.2715 [hep-th]].
}

\lref\SpiridonovZA{
  V.~P.~Spiridonov and G.~S.~Vartanov,
  ``Elliptic Hypergeometry of Supersymmetric Dualities,''
Commun.\ Math.\ Phys.\  {\bf 304}, 797 (2011).
[arXiv:0910.5944 [hep-th]].
}

\lref\BeniniMF{
  F.~Benini, C.~Closset and S.~Cremonesi,
  ``Comments on 3d Seiberg-like dualities,''
JHEP {\bf 1110}, 075 (2011).
[arXiv:1108.5373 [hep-th]].
}

\lref\ClossetVP{
  C.~Closset, T.~T.~Dumitrescu, G.~Festuccia, Z.~Komargodski and N.~Seiberg,
  ``Comments on Chern-Simons Contact Terms in Three Dimensions,''
JHEP {\bf 1209}, 091 (2012).
[arXiv:1206.5218 [hep-th]].
}

\lref\SpiridonovHF{
  V.~P.~Spiridonov and G.~S.~Vartanov,
  ``Elliptic hypergeometry of supersymmetric dualities II. Orthogonal groups, knots, and vortices,''
[arXiv:1107.5788 [hep-th]].
}

\lref\SpiridonovWW{
  V.~P.~Spiridonov and G.~S.~Vartanov,
  ``Elliptic hypergeometric integrals and 't Hooft anomaly matching conditions,''
JHEP {\bf 1206}, 016 (2012).
[arXiv:1203.5677 [hep-th]].
}

\lref\DimoftePY{
  T.~Dimofte, D.~Gaiotto and S.~Gukov,
  ``3-Manifolds and 3d Indices,''
[arXiv:1112.5179 [hep-th]].
}

\lref\KimWB{
  S.~Kim,
  ``The Complete superconformal index for N=6 Chern-Simons theory,''
Nucl.\ Phys.\ B {\bf 821}, 241 (2009), [Erratum-ibid.\ B {\bf 864}, 884 (2012)].
[arXiv:0903.4172 [hep-th]].
}

\lref\WillettGP{
  B.~Willett and I.~Yaakov,
  ``N=2 Dualities and Z Extremization in Three Dimensions,''
[arXiv:1104.0487 [hep-th]].
}

\lref\ImamuraSU{
  Y.~Imamura and S.~Yokoyama,
  ``Index for three dimensional superconformal field theories with general R-charge assignments,''
JHEP {\bf 1104}, 007 (2011).
[arXiv:1101.0557 [hep-th]].
}

\lref\FreedYA{
  D.~S.~Freed, G.~W.~Moore and G.~Segal,
  ``The Uncertainty of Fluxes,''
Commun.\ Math.\ Phys.\  {\bf 271}, 247 (2007).
[hep-th/0605198].
}

\lref\HwangQT{
  C.~Hwang, H.~Kim, K.~-J.~Park and J.~Park,
  ``Index computation for 3d Chern-Simons matter theory: test of Seiberg-like duality,''
JHEP {\bf 1109}, 037 (2011).
[arXiv:1107.4942 [hep-th]].
}

\lref\GreenDA{
  D.~Green, Z.~Komargodski, N.~Seiberg, Y.~Tachikawa and B.~Wecht,
  ``Exactly Marginal Deformations and Global Symmetries,''
JHEP {\bf 1006}, 106 (2010).
[arXiv:1005.3546 [hep-th]].
}

\lref\GaiottoXA{
  D.~Gaiotto, L.~Rastelli and S.~S.~Razamat,
  ``Bootstrapping the superconformal index with surface defects,''
[arXiv:1207.3577 [hep-th]].
}

\lref\IntriligatorID{
  K.~A.~Intriligator and N.~Seiberg,
  ``Duality, monopoles, dyons, confinement and oblique confinement in supersymmetric SO(N(c)) gauge theories,''
Nucl.\ Phys.\ B {\bf 444}, 125 (1995).
[hep-th/9503179].
}

\lref\SeibergNZ{
  N.~Seiberg and E.~Witten,
  ``Gauge dynamics and compactification to three-dimensions,''
In *Saclay 1996, The mathematical beauty of physics* 333-366.
[hep-th/9607163].
}

\lref\KinneyEJ{
  J.~Kinney, J.~M.~Maldacena, S.~Minwalla and S.~Raju,
  ``An Index for 4 dimensional super conformal theories,''
  Commun.\ Math.\ Phys.\  {\bf 275}, 209 (2007).
  [hep-th/0510251].
}

\lref\NakayamaUR{
  Y.~Nakayama,
  ``Index for supergravity on AdS(5) x T**1,1 and conifold gauge theory,''
Nucl.\ Phys.\ B {\bf 755}, 295 (2006).
[hep-th/0602284].
}

\lref\GaddeKB{
  A.~Gadde, E.~Pomoni, L.~Rastelli and S.~S.~Razamat,
  ``S-duality and 2d Topological QFT,''
JHEP {\bf 1003}, 032 (2010).
[arXiv:0910.2225 [hep-th]].
}

\lref\GaddeTE{
  A.~Gadde, L.~Rastelli, S.~S.~Razamat and W.~Yan,
  ``The Superconformal Index of the $E_6$ SCFT,''
JHEP {\bf 1008}, 107 (2010).
[arXiv:1003.4244 [hep-th]].
}

\lref\AharonyCI{
  O.~Aharony and I.~Shamir,
  ``On $O(N_c)$ d=3 N=2 supersymmetric QCD Theories,''
JHEP {\bf 1112}, 043 (2011).
[arXiv:1109.5081 [hep-th]].
}

\lref\GiveonSR{
  A.~Giveon and D.~Kutasov,
  ``Brane dynamics and gauge theory,''
Rev.\ Mod.\ Phys.\  {\bf 71}, 983 (1999).
[hep-th/9802067].
}

\lref\SpiridonovQV{
  V.~P.~Spiridonov and G.~S.~Vartanov,
  ``Superconformal indices of ${\cal N}=4$ SYM field theories,''
Lett.\ Math.\ Phys.\  {\bf 100}, 97 (2012).
[arXiv:1005.4196 [hep-th]].
}
\lref\GaddeUV{
  A.~Gadde, L.~Rastelli, S.~S.~Razamat and W.~Yan,
  ``Gauge Theories and Macdonald Polynomials,''
Commun.\ Math.\ Phys.\  {\bf 319}, 147 (2013).
[arXiv:1110.3740 [hep-th]].
}
\lref\KapustinGH{
  A.~Kapustin,
  ``Seiberg-like duality in three dimensions for orthogonal gauge groups,''
[arXiv:1104.0466 [hep-th]].
}

\lref\orthogpaper{O. Aharony, S. S. Razamat, N.~Seiberg and B.~Willett, 
``3d dualities from 4d dualities for orthogonal groups,''
[arXiv:1307.0511 [hep-th]].
}

\lref\readinglines{
  O.~Aharony, N.~Seiberg and Y.~Tachikawa,
  ``Reading between the lines of four-dimensional gauge theories,''
[arXiv:1305.0318 [hep-th]].
}

\lref\WittenNV{
  E.~Witten,
  ``Supersymmetric index in four-dimensional gauge theories,''
Adv.\ Theor.\ Math.\ Phys.\  {\bf 5}, 841 (2002).
[hep-th/0006010].
}

\lref\KrattenthalerDA{
  C.~Krattenthaler, V.~P.~Spiridonov and G.~S.~Vartanov,
  ``Superconformal indices of three-dimensional theories related by mirror symmetry,''
JHEP {\bf 1106}, 008 (2011).
[arXiv:1103.4075 [hep-th]].
}

\lref\VafaTF{
  C.~Vafa and E.~Witten,
  ``A Strong coupling test of S duality,''
Nucl.\ Phys.\ B {\bf 431}, 3 (1994).
[hep-th/9408074].
}

\lref\WittenWY{
  E.~Witten,
  ``AdS / CFT correspondence and topological field theory,''
JHEP {\bf 9812}, 012 (1998).
[hep-th/9812012].
}

\Title{\vbox{\baselineskip12pt
}}
{\vbox{\centerline{Global Properties of Supersymmetric Theories}
\centerline{}
\centerline{ and the Lens Space}
}}
\centerline{Shlomo S. Razamat and Brian Willett}
\bigskip
\centerline{{\it School of Natural Sciences, Institute for Advanced Study, Princeton, NJ 08540, USA}}
\vskip.1in \vskip.2in \centerline{\bf Abstract}

\noindent 
We compute the supersymmetric 
partition function on $L(r,1)\times \S^1$, the lens space index, for $4d$ gauge theories
 related 
by supersymmetric dualities
and involving non simply-connected groups.
 This computation is sensitive to the 
global properties of the underlying gauge group and to discrete theta angle parameters 
and thus distinguishes
versions of dualities differing by such. We explicitly discuss 
${\cal N}=1$ $so(N_c)$ Seiberg dualities and ${\cal N}=4$ $su(N_c)$ S-dualities.

\vfill

\Date{July 2013}


\newsec{Introduction }

Supersymmetric partition functions are a rare example
of QFT quantities which one can compute exactly. 
Although the set of these quantities is quite limited, they 
provide easily accessible and non trivial  information  
about non-perturbative dynamics of supersymmetric
theories. For example, in the case that the manifold on which the 
supersymmetric partition function is computed is ${\cal M}_3\times \S^1$,
 it has the physical meaning of counting certain protected objects in the field
theory, {\it i.e,} it is an {\it index}.

In this paper we will be interested in a very specific example of such an index 
where the $3d$ manifold ${\cal M}_3$ is taken to be the lens space, $L(r,1)$~\BeniniNC. The $r=1$ case, $L(1,1)=\S^3$, is  the supersymmetric index~\refs{
\RomelsbergerEG,\KinneyEJ}
and was studied in much detail in recent years.  The supersymmetric index encodes 
information about the protected spectrum at the fixed point of the theory, which 
in the case of theories with a known UV weakly coupled description can be 
deduced from group theory considerations alone.  In some cases assuming the global symmetries 
of a theory and some robust information about the dynamics, {\it e.g.,} the behavior at
the cusps of the moduli space, is  sufficient to fix this index completely~\refs{\GaddeUV,\GaiottoXA}.
On the other hand the index has provided a set of highly non-trivial checks 
of conjectured dualities: if several different looking theories are supposed to describe the same 
physics, either for any scale or just in the IR, their supersymmetric indices ought to
agree~\FestucciaWS. Indeed, the indices have been found to agree for many 
of the putative dualities.  For example, the indices of Seiberg IR dualities~\SeibergPQ\
have been checked to agree in~\refs{\DolanQI,\SpiridonovZA},  Gaiotto's ${\cal N}=2$ dualities~\GaiottoWE\
 have
been checked in~\refs{\GaddeKB,\GaddeTE}, and ${\cal N}=4$ S-dualities in~\refs{\GaddeKB,\SpiridonovQV}. Moreover, these indices can be checked to agree with 
computations using dual holographic descriptions of the gauge theories~\refs{\KinneyEJ,\NakayamaUR,\GaddeEN,\EagerHX}.

The computation of the ($\S^3 \times \S^1$) supersymmetric indices of gauge theories 
 is sensitive only to the Lie algebra associated with the gauge group. 
In particular it is insensitive to the global properties of the group, {\it e.g.} 
the indices of  $SU(N)$ and $SU(N)/{{\Z}_N}$ theories are identical.  
However, the global structure of the group 
is believed to be important in supersymmetric dualities. For example, the ${\cal N}=4$ 
S-duality relates a gauge theory with gauge group $G$ with another gauge theory with Langlands
dual gauge group $G^L$:  {\it e.g.}, if the former is $G=SU(N)$ then the latter is $G^L=SU(N)/{{\Z}_N}$. Recently, the authors of~\readinglines\ have thoroughly analyzed the 
implications of different global structures of the gauge groups in known supersymmetric dualities.\foot{In this paper we will assume familiarity with the results of \readinglines, and will not review them in detail here.}
In particular, it was argued in~\readinglines\ that some of the conjectured dualities depend on this global structure in quite a non-trivial way.  For example, when the gauge group is
 not simply connected\foot{
Throughout this paper we will discuss only connected groups. One can generalize the discussion to include also groups which are not connected, such as $O(N)$, but then there will be some new features.
} one may introduce certain discrete theta angles, and these may map in an interesting way under dualities.  The choice of discrete theta angles is closely related to the choice of line operators one includes in the theory, which was also discussed in \GaiottoBE. The details of the dependence of the 4d dualities on the global properties of the gauge group are  
also crucial to understanding aspects of $3d$ dualities following from them~\orthogpaper.
 It would be thus beneficial
to have a simple computational tool which will be sensitive to global structure of the gauge groups and
will give us an indication whether the dualities of~\readinglines\ might be correct.

Here the lens space index with $r>1$ comes onto the stage.  For $r>1$ the manifold $L(r,1)$ has a non-contractible cycle, and as a result it can support non-trivial principal bundles, which are distinguished by certain characteristic classes.  The global structure of the gauge group then enters in which bundles we allow, and how their contribution is weighed in the path integral by the various discrete theta angles.  As a result, the lens indices of theories with different such global structures 
are, in general, different. This gives us a practical tool to probe the dualities discussed in~\readinglines.
The main goal of this paper is thus to test the predictions of~\readinglines\ through
lens space index computations:
in all the cases where we have performed the computation, the lens space index 
is found to be consistent with these dualities.

\

The outline of this paper is as follows.  In section $2$, we review the computation of the lens space index, including new ingredients that must be accounted for when the gauge group is not simply connected.    In section $3$ we consider the simplest example of $G=SU(2)$ and $SO(3)$ in more detail, demonstrating the computation and applying it to test ${\cal N}=4$ S-duality and $SO(3)$ Seiberg duality.  In section $4$ we consider general $so(N_c)$ Seiberg duality, in particular performing a check of the precise mapping between the various versions of these theories explained in \readinglines.  In  section $5$ we consider ${\cal N}=4$ S-duality for $su(N)$ Lie algebras, and perform additional checks of the matching of the indices. We finish in section $6$ by briefly discussing our results.
 An appendix contains additional technical details. 
The manuscript is complemented by a {\it Mathematica} 
notebook  for computing lens indices for the general classes
of theories discussed in the bulk of the 
paper.

\newsec{The lens space index}

In this section we review the ingredients of the matrix model computing the lens space index, {\it i.e.}, the partition function on the space $L(r,1) \times \S^1$, where $L(r,1)$ is the lens space.  The path integral computing this partition function localizes onto  flat connections for the dynamical and background gauge fields.  These can be organized into contributions from various principal bundles, which may enter the partition function with various phases depending on the choice of discrete theta angles, as in \readinglines.
We will see that, unlike the ordinary supersymmetric index, the lens space index is capable of distinguishing theories differing only by the global structure of the gauge group, {\it e.g.}, with different discrete theta angles.

\subsec{The lens space}

To start, let us review some relevant facts about the lens space.  
The lens space $L(p,q)$, for relatively prime positive integers $p$ and $q$, can be defined as a quotient of $\S^3$ 
under the following $\Z_p$ action:

\eqn\Lensact{
 (z_1,\,z_2) \rightarrow ( e^{2 \pi i\,q/p}\, z_1,\, e^{-2 \pi i /p }\, z_2) \,,
}
where we think of $\S^3$ as the subset of $(z_1,z_2) \in {\C}^2$ satisfying $|z_1|^2 + |z_2|^2 = 1$.  This action has no fixed points and so defines a smooth quotient manifold.  We will be interested in the case $q=1$,  and will  adjust our notation to $L(r,1)$ to avoid confusion with some of the standard  index notations introduced below.   In this case the $\Z_r$ action rotates the fibers of a Hopf fibration of $\S^3$.   From now on by the ``lens space'' we will mean the space $L(r,1)$ for some $r \geq 1$.

It will be crucial in this paper that, for $r>1$, the lens space is not simply connected, having fundamental group
$\Z_r$.  We will denote a cycle which generates this group by $\gamma$. 
This is a torsion cycle satisfying $\gamma^r=1$.  A natural way to think about $\gamma$ is
from the quotient description of the lens space. Namely, it can be taken as the image of a path in $\S^3$ connecting two points in the same orbit.  Thus a fiber of the Hopf fibration of $\S^3$
projects to a (contractible) $r$ fold cover of $\gamma$.  The presence of the  cycle $\gamma$ means that the lens space index is sensitive to the global structure of the gauge group, as we will see explicitly in
what follows.

\subsec{Lens space index for free fields}

Let us now review the computation of the lens space index, {\it i.e.}, the supersymmetric partition
function on $L(r,1)\times \S^1$, derived in~\BeniniNC.  One can compute the lens space index in two different ways: first as a partition function
on $L(r,1)\times \S^1$, and second as a counting problem over Hilbert space on $L(r,\,1)$.
Although in what follows the former way of thinking will be more relevant for us, let us first
describe the computation as a counting problem since it is quite intuitive.   We will first review the contributions to the index from free fields, and then in the next sub-section describe how to compute the lens index of a general gauge theory.

We start with the case $r=1$, corresponding to $L(1,1) \cong \S^3$.  To conform to the usual 
notations in the literature, we will refer to the $r=1$ case as the ``supersymmetric index,'' whilst for general $r$ we 
will refer to the $L(r,1)\times \S^1$ supersymmetric partition function as the lens (space) index.
  The supersymmetric index is defined by 
\refs{\RomelsbergerEG,\KinneyEJ}:
\eqn\indexDef{
\II(p,\,q;\,\{u_a\})=\Tr \left[(-1)^F\,e^{-\beta\,\delta}\,p^{j_1+j_2-\frac{R}{2}}\,q^{j_1-j_2-\frac{R}{2}}\,\prod_{a}u_a^{\mu_a}\right]\,.
}  Here $j_1$ and $j_2$ are the Cartan generators of the $SU(2)_1\times SU(2)_2$ isometry of the sphere,
$R$ is the $U(1)_R$ charge, and
the charges  $\mu_a$ correspond to the Cartan ($U(1)$) generators of  global symmetries.  The parameters $p,q$ and $u_a$ on which the index depends are fugacities which couple to these global symmetries.
The chemical potential $\beta$  couples to
\eqn\deltaIn{
\delta\equiv \{{\cal Q},\,{\cal Q}^\dagger\}=E-2j_1+\frac{3}2\,R\,,
} where $E$ is the energy (in conformal theories this is related by the state/operator map to the conformal dimension)
and where we choose ${\cal Q}$ to be the supersymmetry generator with $(j_1,\,j_2)=(-\frac{1}2,\,0)$
and $R=-1$. The index thus is actually independent of $\beta$, since
$j_1\pm j_2-\frac{R}{2}$ and $\mu_a$ commute with ${\cal Q}$. The charges
 $j_{1,2}$ correspond to  momentum along two different Hopf fibers of $\S^3$. Since 
our supercharge ${\cal Q}$ has zero $j_2$ charge, we can consistently quotient by rotations along this Hopf fiber to obtain the lens index for $r>1$, as we will see in a moment.

We now compute the supersymmetric indices of free fields.
The index of a single chiral superfield is given by~\DolanQI
\eqn\indexChi{
{\cal I}^{(R)}_\chi(\{u_a\})=\Gamma((p\,q)^{\frac{R}{2}}\,\prod_a u_a^{\mu_a};p,\,q)\,.
}
Here $R$ is the R-charge of the field and $\mu_a$ are its charges under the global $U(1)$ symmetries. The function on the right-hand side is the elliptic Gamma function
\eqn\ellG{
\Gamma(z;p,\,q)\equiv \prod_{i,j=0}^\infty\frac{1-p^{i+1}q^{j+1}\,z^{-1}}{1-p^iq^j\,z}\,.
} The numerator in the elliptic Gamma function appearing in \indexChi\ comes from 
fermionic modes and the denominator from bosonic ones. The double infinite products corresponds to 
two derivatives contributing to the index (with charges $(j_1,j_2)=(\frac{1}{2},\pm\frac{1}{2})$).

 Note that the index of a chiral field has a natural factorization 
\eqn\factor{
{\cal I}^{(R)}_\chi(z)=
\Gamma((p\,q)^{\frac{R}{2}}z;q,\,pq)\,
\Gamma((p\,q)^{\frac{R}{2}}\,p\,z;p,\,pq)=
\Gamma((p\,q)^{\frac{R}{2}}q\,z;q,\,pq)\,
\Gamma((p\,q)^{\frac{R}{2}}z;p,\,pq)\,.
} This splitting has a simple physical meaning. The derivatives of a 
free scalar field contributing to the index with weight $X$ have the following contribution to the single 
particle index \RomelsbergerEG
\eqn\scalder{
X\,\sum_{i,\,j=0}^\infty p^i\,q^j=X\,\sum_{\ell=0}^\infty\left(p\,q\right)^\ell\left[\sum_{i=0}^\infty (p^i+q^i)-1\right]=\frac{X}{1-p\,q}\left(\frac{1}{1-q}+\frac{p}{1-p}\right).
} The two summands in the last equality come from states with, respectively, non-negative and negative values
of the $j_2$ charge. This is the momentum along the Hopf fiber (preserved by our choice of
the supercharge) of $\S^3$, {\it i.e.}, the momentum around the cycle $\gamma$ after the orbifold projection.

Now let us consider the lens space index for $r>1$.  Recall that the lens space is given by a quotient of $\S^3$ by a rotation of $2\pi/r$ along the Hopf fiber.  Thus to obtain the lens space index, we must first perform a projection onto states which survive under this identification.
The contribution to the  $j_2$ charge from the orbital angular momentum of these states is a multiple of $r$. Thus a scalar contributes
to the single particle index a factor of 
\eqn\Lensscalar{
\frac{X}{1-p\,q}\left(\frac{1}{1-q^r}+\frac{p^r}{1-p^r}\right)\,,
} and the index of a chiral multiplet on the $L(r,1)\times \S^1$ is then
\eqn\factor{
{\cal I}^{(R)}_\chi(z)=
\Gamma((p\,q)^{\frac{R}{2}}z;q^r,\,pq)\,
\Gamma((p\,q)^{\frac{R}{2}}\,p^r\,z;p^r,\,pq)=
\Gamma((p\,q)^{\frac{R}{2}}q^r z;q^r,\,pq)\,
\Gamma((p\,q)^{\frac{R}{2}}\,z;p^r,\,pq)\,.
} 

In addition to counting  states which are invariant under the orbifold projection, one can also consider states which transform by a non-trivial phase.\foot{When the symmetry in question is gauged, as in the next subsection, we must also include these states in the index, analogous to the
  ``twisted sectors'' one must include when performing an orbifold projection.}  Namely, we can consider a chiral field with unit charge under a $U(1)_u$ symmetry for which, in addition to the 
fugacity $u$, we also turn on a non trivial holonomy around the non-contractible cycle $\gamma$.  Specifically, in taking the field around $\gamma$, it acquires a phase $e^{\frac{2 \pi i m}{r}}$,  ($0\leq m<r$). Thus,
in presence of such a holonomy the projection on the modes of the scalar field is now that the 
charge $j_2$ has to be equal $m\,({\rm mod}\,r)$. Since the index 
of a chiral field gets contribution from a scalar and from the fermion in the complex
 conjugate multiplet the projection on fermions contributing to the index  is to states with 
$j_2$ charge equal $-m\,({\rm mod}\,r)$.\foot{Note that this  was  not implemented
in~\BeniniNC\ and the expressions for the free fields there are slightly different from the ones appearing here.}  All in all, the index of a chiral field in presence of a holonomy becomes\foot{
The factorization into a product of two elliptic Gamma functions here is akin to the factorization into 
blocks of the $3d$ index~\refs{\PasquettiFJ,\BeemMB} (see also \KrattenthalerDA) as can be seen by taking $r\to\infty$ limit.
}
\eqn\holUone{
{\cal I}^{(R)}_\chi(m,\,u)= {\cal I}_0(m,u)\;
\Gamma((p\,q)^{\frac{R}{2}}\,q^{r-m}\,u;q^r,\,pq)\,
\Gamma((p\,q)^{\frac{R}{2}}\,p^m\,u;p^r,\,pq)\,.
} Note that the holonomy breaks the symmetry between $p$ and $q$ which we had until now
since for general values of $m$ the projection on non-negatively and negatively charged states is different. The factor ${\cal I}_0(m,u)$ is the zero point energy which has to be introduced 
in presence of non-trivial $m$ since the vacuum here acquires a charge. 
If the symmetry for which we turn on a non trivial holonomy has no anomalies (global or local),
 the contribution to the zero point energy of a single chiral is given by~\BeniniNC,
\eqn\zp{
{\cal I}_0(m,u)=\left((p\,q)^{\frac{1-R}{2}}\,u^{-1}\right)^{\frac{m\,(r-m)}{2r}}\left(
\frac{p}{q}\right)^{\frac{m(r-m)(r-2m)}{12r}}\,.
} In cases when the symmetry for which we turn on the holonomy has an anomaly, more care is 
to be exercised. In this paper we will not consider holonomies for global symmetries, and any local
symmetry is anomaly free, so the above expression will be the relevant one for us.

One can repeat the  same analysis for the lens space index of a free vector field with the result being
\eqn\vectLens{
{\cal I}_V(m,\,u)= \frac{{\cal I}^{V}_0(m,u)}
{(1-u^{-1})^{\delta_{m,0}}\,
\Gamma(q^{m}\,u^{-1};q^{r},\,pq)\,
\Gamma(p^{r-m}\,u^{-1};p^r,\,pq)
}\,.
}  Here $u$ is a fugacity for a $U(1)$ component of the Cartan.
The zero point energy for the vector field is given by~\BeniniNC, 
\eqn\zpV{
{\cal I}^V_0(m,u)=\left((p\,q)^{\frac{1}{2}}u^{-1}\right)^{-\frac{m(r-m)}{2r}}\left(
\frac{q}{p}\right)^{\frac{m(r-m)(r-2m)}{12r}}\,.
}
From the contributions of the vector and the matter fields we can build the lens space index of any gauge theory.

\subsec{Lens space index of a gauge theory}

Above we constructed the lens space index for a free field by starting with the result for $r=1$ and performing a projection in the Hilbert space of the theory on $\S^3 \times \S^1$ onto states invariant under the $\Z_r$ action defining $L(r,1)$, allowing also ``twisted'' states which transform by a phase under this action.  One could also obtain this result in the path integral language by localization.  In that case, one finds that the partition function on $L(r,1)\times \S^1$ for a gauge theory with gauge group $G$ localizes onto flat connections, where the ``twisted'' states correspond to the connections with non-trivial holonomies around the cycle $\gamma$ ~\BeniniNC.  The contribution from a given flat connection is given by the index for the free fields of the theory in this background, which was presented above.  Let us now discuss how one performs the sum over the contributions from all of these flat connections to obtain the lens index of a gauge theory.  Throughout this paper we assume that $G$ is connected.

The flat connections are labeled by two holonomies around the two cycles of $L(r,1)\times \S^1$: the lens
cycle $\gamma$ and the temporal $\S^1$.  We label the two holonomies  by $g$ and $h$, respectively, which are elements of the gauge group $G$, considered up to simultaneous conjugation.  In order that these determine a homomorphism from the fundamental group of $L(r,1) \times \S^1$ to $G$, we must have:

\eqn\holoeqn{
g^r = 1, \qquad\qquad g\,h\, g^{-1} h^{-1} = 1\,. }
Considering $g$ and $h$ up to simultaneous conjugation by an element 
of $G$, this leaves a discrete set of choices for $g$.  Namely, we may always assume $g$ is in a chosen maximal torus of $G$, and write:
\eqn\gdef{
g \rightarrow \{ e^{2 \pi i m_i/r} \}, \;\;\; i=1,...,r_G , }
where $r_G$ is the rank of $G$, and we expand $g$ in the generators of the maximal torus.  In general there will still be a continuous set of choices for $h$.  We will denote the contribution to the index from flat connections defined by $g$ and $h$ 
as ${\cal I}_{g,\,h}$; it is given by a product of the free field indices of the previous sub-section.

\midinsert\bigskip{\vbox{{\epsfxsize=3.in
        \nobreak
    \centerline{\epsfbox{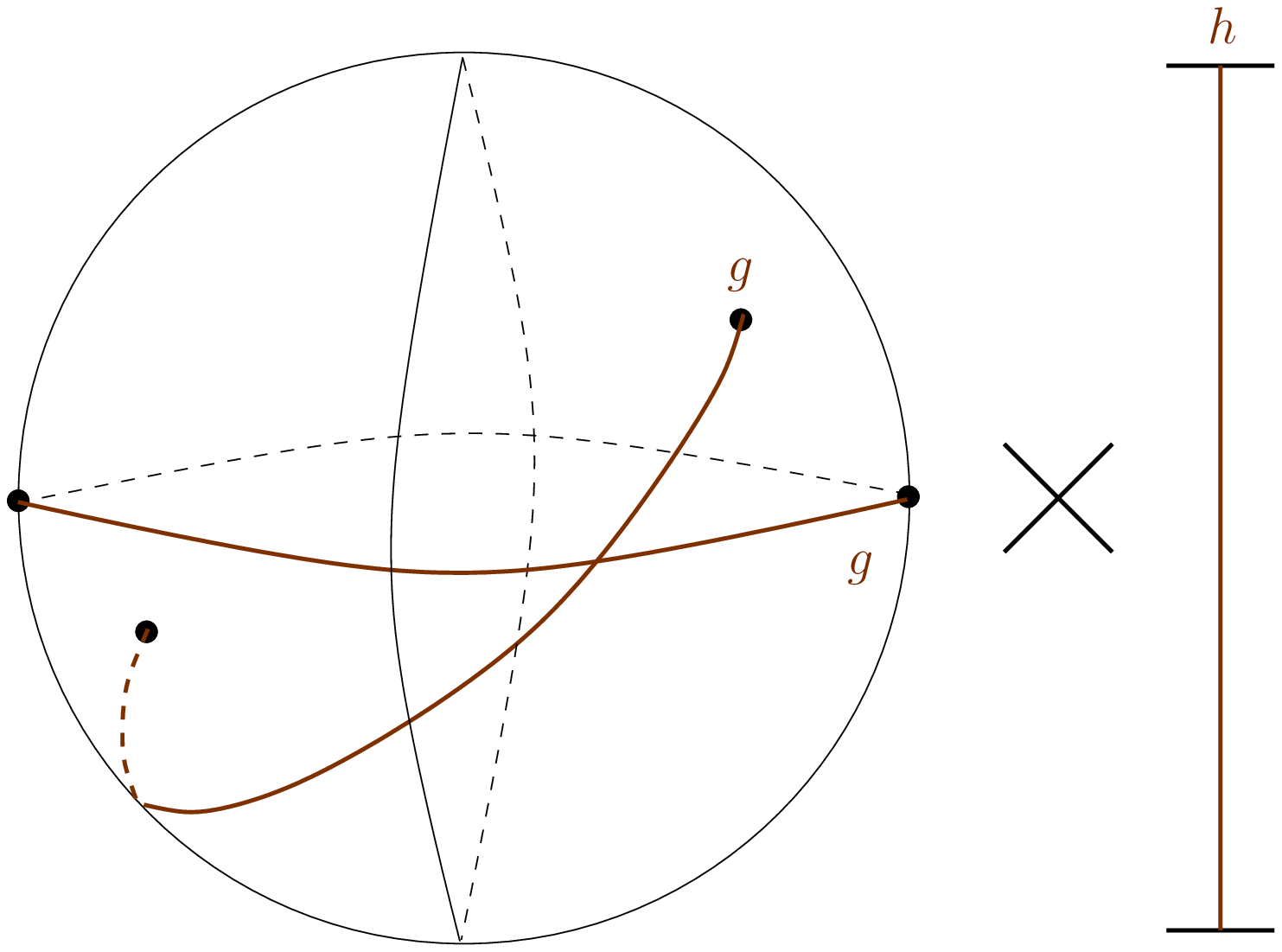}}
        \nobreak\bigskip
    {\raggedright\it \vbox{
{\bf Fig 1.}
{\it Depicted here is the space $L(2,1)\times \S^1\sim \R \P^3\times \S^1$.
$L(2,1)$  is a solid ball with the antipodal points on the boundary identified, and
$\S^1$ is a segment with identified boundaries.
 The group elements $g$ and $h$
are the holonomies around the two cycles.
 In 
bold (brown) are curves wrapping the non-contractible cycles. Wrapping twice around the cycle
of $L(2,1)$ the curve becomes contractable, and thats why here $g^2=1$.
The two cycles commute and thus $g$ commutes with $h$. 
}}}}}}
\bigskip\endinsert

In case that the group $G$ is simply connected any solution $(g,h)$ to the second condition in~\holoeqn\ is equivalent under simultaneous conjugation to a solution where $g$ and $h$ are both in the maximal torus of $G$, so without loss we may consider such solutions.  Then the holonomies $g$ can be parameterized by a set of integers $\{m_i\}, i=1,...,r_G$, as above, and the index takes the following form
\eqn\indexgaugeone{
{\cal I}=\sum_{g,\, h}{\cal I}_{g,\,h} \equiv \sum_{\{m_i\}}\,{\cal I}_{\{m_i\}}\,,
} where
\eqn\gaugelens{\eqalign{
&{\cal I}_{\{m_i\}}=\frac{{\cal I_V}(0,1)^{r_G}}{|W_{\{m_i\}}|}
\oint \prod_{\ell=1 }^{r_G}\frac{dz_\ell}{2\pi i z_\ell} \Delta_{\{m_i\}}(z_i)\prod_{\alpha\in R_+}
{\cal I}_V(\pm\alpha(\{m_i\}),\, e^{\pm\alpha(\epsilon)})\;\times\cr
&\qquad\qquad \qquad\qquad \qquad \prod_{\ell=1}^{N_\chi}
{\cal I}_\chi^{(R_\ell)}(\rho_\ell(\{m_i\}), e^{\rho_\ell(\epsilon)}\,\prod_a u_a^{\mu^{(\ell)}_a})\,.
}} We use the usual short-hand notation $f(x_\pm)=f(x_+)\,f(x_-)$.
The integral here is over the maximal torus of the group and corresponds to integrating over the 
choices of $h$ discussed above. 
The factor $|W_{\{m_i\}}|$ is the order of the Weyl group preserved by holonomy $\{m_i\}$ and  $\Delta_{\{m_i\}}(z_i)$
is the Haar measure of the un-broken group.  The sum should be taken only over Weyl-inequivalent choices of the $\{m_i\}$.  
In addition, we have diagonalized the action of $h$ on the chiral and vector multiplets by decomposing them into the weight spaces of the relevant representations.  For the vectors, we have defined $R_+$ to be the set of positive roots; these are linear functionals acting on the Cartan of the Lie algebra, for which we take a basis $\epsilon_i$, and we define $z_i=\exp(\epsilon_i)$. We have also assumed there are $N_\chi$ chiral fields
with R-charges $R_\ell$, weights $\rho_\ell(\epsilon)$ under the gauge group, and charges $\mu^{(\ell)}_a$ under 
the flavor group. 

In the case of non-simply connected groups, it is no longer the case, in general, that any two elements $g$ and $h$ which commute can be simultaneously conjugated to maximal torus~\WittenNV.\foot{
We thank E.~Witten for illuminating discussions on this issue.
}  The problem can be phrased as follows.  Let $\widetilde{G}$ be the simply connected covering group of $G$, and let $H$ be a subgroup of the center of $\widetilde{G}$ such that $G \cong \widetilde{G}/H$.  Then we can pick lifts $\widetilde{g}$ and $\widetilde{h}$ in $\widetilde{G}$ of $g$ and $h$, and define:

\eqn\NUdef{ \nu = \widetilde{g}\; \widetilde{h} \; \widetilde{g}^{-1}\, \widetilde{h}^{-1} \,.}
Since this projects to the identity in $G$, $\nu$ is an element of $H$.  Moreover, since $H$ is a subgroup of the center of $\widetilde{G}$, we see that changing the lifts by an element of $H$ does not affect $\nu$.  Those commuting pairs $(g,h)$ which can be simultaneously conjugated to the maximal torus of $G$ correspond to $\nu=1$, but there may also be solutions for non-trivial $\nu$, and  these cannot be conjugated to the maximal torus.  We will refer to pairs $(g,h)$ corresponding to non-trivial $\nu$ as ``almost commuting pairs'' following the notations of~\refs{\almost,\WittenNV,\SchweigertTG}.

As an example of a solution with non-trivial $\nu$, take $G=SO(3)$, so that $\widetilde{G}=SU(2)$, and consider:

\eqn\sothreeAC{ \widetilde{g} = \pmatrix{i & 0 \cr 0 & -i }, \qquad
\qquad \widetilde{h} =  \pmatrix{ 0 & 1 \cr -1 & 0 } \,.
}
Then one can see that $\widetilde{g}\, \widetilde{h} = - \widetilde{h}\, \widetilde{g}$, so that these elements do not commute in $SU(2)$, but their images in $SO(3)\cong SU(2)/\{\pm 1\}$ will commute.  In this particular example, as we will show in the next section, any solution with $\nu=-1$ can be conjugated to the pair of elements above, but in general there will be a moduli space of gauge-inequivalent solutions for each choice of $\nu$.  We will describe the solutions to \NUdef\ explicitly for the cases $G=SO(N)$ and $G=SU(N)/\Z_d$ as we consider them in the following sections.  See 
\refs{\SchweigertTG,\almost}\ for a discussion of the solutions of this equation for an arbitrary Lie group $G$.

Next we must describe the contribution to the lens index from a flat connection with almost commuting holonomies $g$ and $h$.  Recall that above the contributions of different fields were given by decomposing the fields into components corresponding to the weight spaces of the relevant representations of the gauge group.  In the present case, since the holonomies $g$ and $h$ do not both lie in the maximal torus, their action on the usual weight basis will not be diagonal.  However, provided the relevant representations are  good representations of $G$ (as opposed to $\widetilde{G}$), there will be some basis in which the action of both $g$ and $h$ is diagonal.

Let us return to the example of $G=SO(3)$, and consider a chiral multiplet in the spin-$1$ representation.  In this representation, the elements in \sothreeAC\ commute in $G$, and so can be represented by diagonal matrices.  To do this, rather than expanding in the weight space basis, let us expand in the basis natural when $SO(3)$ acts on $\R^3$,\foot{Specifically, if we write this $SO(3)$ basis as $(e_x,\,e_y,\,e_z)$, spanning $\R^3$, then the weight basis would have been $e_x \pm i e_y$ and $e_z$.} in which case we find 

\eqn\SOmatrices{R(g)=\pmatrix{-1 & 0 & 0 \cr 0 & -1 & 0 \cr 0 & 0 & 1} , \qquad\qquad
 R(h) = \pmatrix{-1 & 0 & 0 \cr 0 & 1 & 0 \cr 0 & 0 & -1} \,.}
Since we must also impose that $g^r=1$, we see these holonomies can only occur for $r$ even.  Then if we exapand the components of the chiral field of R-charge $R$ in this basis, we find its contribution can be written as

\eqn\chiralAlmost{ 
{\cal I}_\chi^{(R)}\left(\frac{r}2,1\right)\;\;
{\cal I}_\chi^{(R)}(0,-1)\;\;
{\cal I}_\chi^{(R)}\left(\frac{r}2,-1\right)\,.}
Here ${\cal I}_\chi^{(R)}(m,\,z)$ was defined in~\holUone, and we have used the fact that the $-1$ eigenvalues of $R(g)$ correspond to $m=r/2$. Since $R(g)=\{e^{\frac{2\pi i m_i}{r}}\}$
and $R(h)=\{z_1,\,z_2,\,z_3\}$, 
 the holonomy $R(g)$ in \SOmatrices\ implies $\{m_i\}=\{\frac{r}{2},\,\frac{r}{2},\,0\}$, whereas 
$R(h)$ in \SOmatrices\ implies that $\{z_i\}=\{-1,\,1,\,-1\}$.
Note that there is no remaining integral over $z$ to perform in this case, since all other almost commuting holonomies are gauge equivalent to this one.  In general, for each $\nu$, there will be a discrete sum over $g$ and integral over $h$ corresponding to gauge-inequivalent solutions to \NUdef.

\subsec{Flat connections, holonomies, and principal bundles}

In the previous sub-section we introduced the discrete parameter, $\nu$, labeling classes of solutions to \NUdef.  Here we will see that $\nu$ turns out to contain information about the topological type of the principal $G$-bundle over $L(r,1) \times \S^1$ on which this flat connection lives.  We will introduce an additional parameter, $\mu$, such that the pair $(\mu,\nu)$ completely specifies the (flat) bundle over $L(r,1) \times \S^1$.   We can then organize the computation of the partition function into a sum over contributions $Z_{\mu,\nu}$ from each of these $G$-bundles.  To do this we must describe how the  holonomies $g$ and $h$ are related to the topological data classifying the bundle.

As a warm up, let us classify the principal bundles over $L(r,1)$.  The principal $G$-bundles over any three manifold ${\cal M}_3$ are classified by an element of $H^2({\cal M}_3,\pi_1(G))$, which can be thought of as measuring the obstruction to extending a trivialization of the bundle from the $1$-skeleton of ${\cal M}_3$ to the $2$-skeleton; since $\pi_2(G)=0$ there is then no further obstruction to extending it to all of ${\cal M}_3$.  In particular, for a simply connected group there is only the trivial bundle over ${\cal M}_3$.  In the case $G=SO(N)$ this characteristic class of the bundle is known as the second Stiefel-Whitney class, and we will often refer to this class as a Stiefel-Whitney class for general $G$.

Now consider the lens space, $L(r,1)$.  We will need to compute its second cohomology with coefficients in an arbitrary abelian group $A$.  For this purpose, we note it has a cell structure with one $1$-cell and one $2$-cell, where the boundary of the latter is attached to the former by the $r$-fold covering map.  Thus the relevant maps between chain groups fit into the following diagram:

\eqn\lensCC{\matrix{
&&C_2&&C_1&&\cr
&&\cong&&\cong&&\cr
0 & \rightarrow & \Z &  {\buildrel{r}\over\longrightarrow\;} & \Z & \rightarrow& 0 \,.\cr
}}
In particular, the first homology group is $\Z_r$.  Dualizing this with $ {\rm Hom}(\;\_\;,A)$, we obtain maps:

\eqn\lensCCC{\matrix{
&&C^2&&C^1&&\cr
&&\cong&&\cong&&\cr
0 & \rightarrow &{\rm Hom}(\Z,A) &  {\buildrel{r^*}\over\longleftarrow\;} & {\rm Hom}(\Z,A) & \rightarrow& 0 \,.\cr
}}
where now the map $r^*$ sends an element $a \in A$ to $a^r$ and ${\rm Hom}(\Z,A)\cong A$.  For simplicity we will subsequently also denote this map by $r$.  Thus we find:

\eqn\lenssecondhomol{ H^2(L(r,1),A) = A/ A^r\,.}
When $A=\pi_1(G)$, we expect that there is a principal $G$-bundle for every choice of an element in this group.  Moreover, each such bundle supports a flat connection, and there is a 
map from the holonomy of this connection around $\gamma$, which is some element of $G$ we will denote by $g$, to the element $w \in H^2(L(r,1),\pi_1(G))$ labeling the bundle.  To determine this correspondence, let us take 
 $G \cong \widetilde G/H$ as before, with $H \cong \pi_1(G)$.  Then $g^r=1$, which implies that $\widetilde g^r$ is an element of $H$.  However, $\tilde g$ was not uniquely determined, rather we are free to multiply it by any element of $H$, which shifts $\tilde g^r$ by an element of $H^r$.  Thus, given $g$, we can uniquely define $\mu\in H/H^r $ as 

\eqn\mudef{
\mu\; \equiv\; [\tilde g^r] \in H/H^r\,.
}
We claim this corresponds precisely to the characteristic class of the bundle on $L(r,1)$ in $H^2(L(r,1),H)$.

Let us move on to $L(r,1) \times \S^1$.  Since $\pi_3(G)$ is non-trivial, there is an additional invariant of a principal $G$-bundle over a four-manifold, ${\cal M}_4$, namely, the instanton number, which may be non-trivial even when we consider the simply connected group $\tilde{G}$.  However, there are still bundles which are $G$-bundles but do not lift to $\tilde{G}$-bundles, and these are again labeled by a class in the second cohomology.  By a similar computation as above,\foot{Namely, the chain complex splits, with another copy of the complex in \lensCC\ shifted one dimension higher, arising from taking the product with the $\S^1$ cycle.  Dualizing this complex gives the second factor in (2.25).} we find

\eqn\lensindsecondhomol{H^2( L(r,1) \times \S^1 , A ) = A/A^r \oplus {\rm ker} \; r\,}
where the second factor is the kernel of the map $r$ from $A$ to $A$ which takes an element to its $r$th power.

Again we can label the bundles by the flat connections they admit.\foot{In particular, we  specialize to zero instanton number.  Some of the classes in $H^2(L(r,1) \times S^1)$ only arise for bundles with certain fractional instanton numbers, and so for these classes we will not find any contribution to the index.}  Defining $\mu$ as above, it corresponds to the first factor in $H^2(L(r,1) \times \S^1,\,H)$.  We claim that the element $\nu\in H$ we defined in \NUdef\ corresponds to the second factor in $H^2(L(r,1) \times \S^1,H)$.  To see it is really in the kernel of the map $r$,
note that, by rearranging~\NUdef, we find:

\eqn\Nuspace{ 
\nu\, \widetilde{g}^{-1} = \widetilde{h}\, \widetilde{g}^{-1} \widetilde{h}^{-1} \qquad
\Rightarrow\qquad  \nu^r\, \widetilde{g}^{-r} = \nu^r\, \mu^{-1} = \widetilde{h}\, \mu^{-1} \widetilde{h}^{-1} = \mu^{-1}  \qquad \Rightarrow\qquad  \nu^r = 1\,,}
where we have used the fact that $\mu$ and $\nu$ lie in the center of $\widetilde G$.  Thus $\nu \in {\rm ker}\; r$, and so it is natural to postulate the following correspondence between the holonomies $g$ and $h$ and the characteristic class of the bundle as:
 
\eqn\ghmunumap{
(g,h) \rightarrow (\mu, \nu ) \in H^2(L(r,1) \times \S^1,\pi_1(G)) \,,
}
where $\mu$ and $\nu$ are defined through \mudef\ and \NUdef.  In the appendix this correspondence is shown explicitly in the case of $SO(N)$ bundles over $L(2,1) \cong \R \P^3$ by reducing the associated vector bundles to direct sums of line bundles.  

With this in mind, let us now  organize the computation of the index to be a sum over 
 sectors labeled by $\mu$ and $\nu$, which we denote $Z_{\mu,\nu}$. We consider gauge theory with gauge group  $G\cong \widetilde G/H$. 
First consider the index of the simply connected cover $\widetilde G$, which is given as in~\indexgaugeone\ by

\eqn\gaugetwo{ {\cal I}_{\widetilde G} = \sum_{\widetilde g,\widetilde h \in G;\; {\widetilde g}^r=\widetilde g\widetilde h{\widetilde g}^{-1} {\widetilde h}^{-1}=1} {\cal I}_{{\widetilde g},{\widetilde h}} \,.}
 In this expression we implicitly have chosen some particular 
gauge, and in particular the sum is over gauge-inequivalent choices of ${\widetilde g}$ and $\widetilde h$. Computing the index of a theory with gauge group $G$ we have  less gauge redundancy 
but should allow more holonomies,

\eqn\liftItilde{ {\cal I}_{G_0} = \frac{1}{|H|} \sum_{\widetilde g,\widetilde h \in \widetilde{G};\, 
{\widetilde g}^r, {\widetilde g}{\widetilde h}{\widetilde g}^{-1}{\widetilde h}^{-1} \in H} {\cal I}_{\widetilde g,\widetilde h}\,.}
Specifically, the volume of the gauge group $\widetilde G$ is larger than the the volume of $G$ by $|H|$, and since $H \times H$ acts freely on $(\widetilde{g},\widetilde{h})$, for each choice of $g$
and $h$ in $G$ there are $|H|^2$ different lifts to $\widetilde G$, hence the $1/|H|$ overall factor 
above.
We have introduced  label $0$ for the gauge group, as in~\readinglines, denoting that all the sectors are summed with the same weight. 
One can view this equation as an explicit definition of~\indexgaugeone\ for non-simply connected 
group. Now we split this into sectors by specifying $\mu \in H/H^r$ and $\nu \in {\rm ker}\; r$, namely,

\eqn\zmunudef{ 
Z_{\mu,\nu} \equiv  \sum_{
\widetilde g,\widetilde h \in \widetilde{G},\;[{\widetilde g}^r]=\mu, \; {\widetilde g}{\widetilde h}{\widetilde g}^{-1}{\widetilde h}^{-1} =\nu } {\cal I}_{\widetilde g, \widetilde h}\,,}
where $[\widetilde g^r]$ denotes the equivalence class of $\widetilde g^r$ in $H/H^r$.  Then we can write,

\eqn\indzmunu{{\cal I}_{G_0} = \frac1{|H|}\sum_{\mu \in H/H^r, \nu \in {\rm ker}\; r} Z_{\mu,\nu}\,.}
In following sections we will compute explicitly $Z_{\mu,\nu}$ for a variety of examples.

We have thus organized the computation of the index into a sum of contributions  $Z_{\mu,\nu}$ from the
different  bundles. Each such bundle has a Stiefel-Whitney class $(\mu,\nu)$ associated with it.
 It is important to emphasize, however, that the path integral contribution from a single bundle is not a quantity that is defined by  local considerations.  In a physical path integral, one must sum over all the bundles.  In ${\cal I}_{G_0}$ above we have done this while giving equal weight to all the bundles, but more generally we may weigh them with different phases if such phases arise from some local term
which can be added to the action.  An illustrative example of this is the instanton number: here we are not free to specify a bundle with a chosen instanton number, however we can insert the local term $\frac{i\,\theta}{16 \pi^2}\, F \wedge F$, which weighs the various bundles with instanton number $n$ by a phase $e^{i n \theta}$.
In the present case we are not sensitive to the theta angle of the theory, since we only get contributions from flat connections.  However, one can still include discrete theta angles, as in~\readinglines.  These are local terms that depend on the Stiefel-Whitney class of the bundle, and can evaluate to non-zero phases  even on flat bundles, as the ones above.  Thus, depending on the choice of these discrete theta angles, the different sectors $Z_{\mu,\nu}$ may enter into the partition function multiplied by different phases.  Specifically,  given a discrete theta angle which assigns a rational number $c(w_2)$ to a bundle with the second Stiefel-Whitney class $w_2$, we can define,

\eqn\gcdef{
{\cal I}_{G_c} = \frac{1}{|H|} \sum_{\mu,\nu} e^{i c(\mu,\nu)} Z_{\mu,\nu}\,.
}
To take an example which is central in this paper, it was argued in \readinglines\ that theories with gauge group $SO(N)$ admit a discrete theta angle proportional to $P_2(w_2)$. Here \eqn\pont{P_2:H^2({\cal M}_4,\Z_2) \rightarrow H^4 ({\cal M}_4,\Z_4)\,,} is the Pontryagin square operator.
Since  $\pi_1(SO(N))\cong \Z_2$, we see from \lensindsecondhomol\ that for $L(r,1) \times \S^1$ we only have non-trivial $\Z_2$ cohomology for $r$ even, and so we will specialize to this case in order to be sensitive to this discrete theta angle.  Then $H^2(L(r,1) \times \S^1,\Z_2) = \Z_2 \oplus \Z_2$, and an element of the second cohomology can be labeled by $(\mu,\nu)$ as above.  We will see in the next section that:

\eqn\pontsq{ 
r=0 \; ({\rm mod}\;  4)\; : \qquad 
P_2\left(\mu,\nu\right) = \cases{
2 & $\mu=-1,\nu=-1$ \cr
0 & ${\rm else}$ }\,,
}

$$
r=2 \; ({\rm mod}\;  4)\; : \qquad 
P_2\left(\mu,\nu\right) = \cases{
2 & $\mu=1,\nu=-1$ \cr
0 & ${\rm else}$ }\,.
$$
where on the right-hand side the integer is understood to be multiplying the generator of $H^4(L(r,1) \times \S^1,\Z_4) \cong \Z_4$.  Then when the gauge group is $SO(N)$ we may choose either 
not to insert the discrete theta angle and obtain the so called $SO(N)_+$ theory, or to insert it with a non-zero coefficient, which gives the  $SO(N)_-$ theory~\readinglines.  To compute the lens index of the $SO(N)_-$  theory, we must then weigh with a minus sign the contribution $Z_{\mu,\nu}$ corresponding to the class
$(\mu,\nu)$ which has a non-zero Pontryagin square.

Before moving on let us note that, for the purpose of computation, one can always, without loss, specialize to $\hat{G}=\tilde{G}/\hat{H}$, where $\hat{H}$ is the maximal subgroup of the center of $\widetilde{G}$ which acts trivially on the matter.  This is because the sectors for any other gauge group $G$ we could choose can be built out of the sectors $Z_{\mu,\nu}$ for $\hat{G}$.  Namely, if we write $G = \tilde{G}/H$, then $H$ is subgroup of $\hat{H}$, so there are maps

\eqn\idefs{
 i: H \hookrightarrow \hat{H}, \qquad\qquad \hat{i}: H/H^r \rightarrow \hat{H}/\hat{H}^r \,,}
where the former is the inclusion map, and in the latter we use the fact that this maps $H^r$ into $\hat{H}^r$ to pass the map through to the quotient groups (however it is important to emphasize that $H/H^r$ is not, in general, a subgroup of $\hat{H}/\hat{H}^r$).  We can use these maps to map a choice of $(\mu,\nu)$ for $G$ to a corresponding choice for $\hat{G}$.  Then we find:
\eqn\gfromgo{ Z_{\mu,\nu}^G = \; Z_{\hat{i}(\mu),i(\nu)}^{\hat{G}} }

Let us now turn to computing the lens space index in a simple example which illustrates explicitly some of the points discussed here.

\newsec{$su(2)$ gauge theory}

Consider a gauge theory with lie algebra $su(2)$.  The simply connected group based on this algebra, $SU(2)$, has a center $\Z_2$, which acts trivially on the integer spin representations but non-trivially on the half-integer spin ones.  Thus if the matter of the theory only comes in representations of integer spin, one may instead choose to consider the group $SU(2)/\Z_2$.  This gives a different theory, since now one must include additional principal bundles: namely those which are $SU(2)/\Z_2$ bundles but do not lift to $SU(2)$ bundles.  

As described in the previous section the distinction between the various bundles on $L(r,1) \times \S^1$ enters the index in certain algebraic properties of the holonomies $g$ and $h$ around the non-trivial cycles
which we labeled by $\mu$ and $\nu$ in the previous section.  In this case, $H=\Z_2$, and the parameters $\mu,\nu$  take values in:

\eqn\munusutwo{\eqalign{
& \mu \in H/H^r \cong \cases{ \Z_2   & $r$ even;\cr 0 , & $r$ odd}\,,\cr
&\cr
& \nu \in {\rm ker}\;r \cong \cases{ \Z_2   & $r$ even;\cr 0 , & $r$ odd }\,.
}}
In particular, for $r$ odd the lens space index is not sensitive to the difference between $SU(2)$ and $SO(3)$.  Thus we focus on the case of $r$ even.  Then we see the holonomies $g,\,h$ can be grouped into four classes determined by $\mu,\,\nu\in \{1,-1\}$.

We now consider the problem of finding pairs of $SU(2)$ matrices satisfying the various conditions \mudef \ and \NUdef.  We pick $\widetilde{g}$ to lie in the maximal torus of $SU(2)$.  Then we can take

\eqn\sutwog{ \widetilde{g} = \pmatrix{ e^{\pi i m/r}& 0 \cr 0 & e^{-\pi i m/r} }\,. }
The choices with $m$ even correspond to $\mu=1$, while those with $m$ odd correspond to $\mu=-1$.
  Note that $m$ is only determined modulo $r$ by $\widetilde g$,
 however its parity is well-defined since $r$ is even.
Next we must find the allowed $\widetilde{h}$.  Let us write the most general choice of $\widetilde{h}$ as

\eqn\tildehsutwo{ \tilde{h} =\pmatrix{ h_{11} & h_{12} \cr h_{21} & h_{22} }\,.  }
Let us first look for solutions with $\nu=1$, {\it i.e.}, so that $\widetilde g$ and $\widetilde h$ commute.  Then the condition that $\widetilde{h}$ commute with our choice of $\widetilde{g}$ leaves $h_{11}$ and $h_{22}$ arbitrary, but forces:

\eqn\hssutwo{ h_{12}\,( e^{\pi i m/r} - e^{-\pi i m/r}) = h_{21}\, ( e^{\pi i m/r} - e^{-\pi i m/r})  = 0\,. }
Thus either $h_{12}=h_{21}=0$, in which case $\widetilde{h}$ is in the maximal torus, or $e^{2 \pi i m/r}=1$, {\it i.e.}, $\widetilde{g}=\pm 1$.  But in the latter case there is now a larger residual gauge symmetry which allows us to rotate $\tilde{h}$ into the maximal torus, and so we may always assume $\widetilde{h}$ is diagonal.  We write it as

\eqn\tldehsutwoB{ \widetilde{h} = \pmatrix{ e^{i a }& 0 \cr 0 & e^{-i a } }\,.}
Next we look for solutions with $\nu=-1$, {\it i.e.}, such that $\widetilde{h}$  anticommutes with our choice of $\widetilde{g}$.  This time we are forced to set $h_{11}=h_{22}=0$, as well as

\eqn\anticomhsutwo{ h_{12}\,( e^{\pi i m/r} + e^{-\pi i m/r}) = h_{21}\, ( e^{\pi i m/r} + e^{-\pi i m/r})  = 0\,. }
Since $\widetilde{h}$ is an $SU(2)$ matrix, we cannot also set $h_{12}$ and $h_{21}$ to zero, and so we must impose $e^{2 \pi im/r}=-1$, {\it i.e.}

\eqn\tildegsutoB{ \widetilde{g} = \pm \pmatrix{i & 0 \cr 0 & -i }\,. }
Now any choices of $h_{12}$ and $h_{21}$ will lead to an anti-commuting matrix.  However, for this matrix to lie in $SU(2)$ we must impose $h_{12}\, h_{21}=-1$.  Moreover, we are still free to conjugate this matrix by any element of the maximal torus, since this preserves $\widetilde{g}$, and all allowed choices of $\widetilde{h}$ are related by such a conjugation.  Thus there is a single point in the space of flat connections on this bundle, and we may take, {\it e.g.}

\eqn\tildehsutwoC{ \widetilde{h} =  \pmatrix{ 0 & 1 \cr -1 & 0 }\,. }

Note $\tilde g^r =1$ for $r=0 \;({\rm mod}\;4)$ and $\widetilde g^r = -1$ for $r=2 \;({\rm mod}\;4)$.  Thus the sector $(\mu,\,\nu)=(-1,-1)$ is missing for $r =0\;({\rm mod}\;4)$ while for
$r=2 \;({\rm mod}\;4)$ the sector $(\mu,\,\nu)=(1,-1)$ is missing.  This is related to the mod $4$ behavior in \pontsq, specifically, since $P_2(w_2)/4$ for $SO(3)$ is equal to the instanton number, modulo $1$, we see that the choices for $(\mu,\nu)$ which do not correspond to flat connections are precisely those for which $P_2(w_2)\neq 0$, and so also the instanton number, is non-zero.  Thus these bundles do not contribute to the index, and for the $SO(3)$ lens index there are $3$ non-trivial contributions to the partition function, from the sectors $(\mu,\,\nu)=(1,1),\, (-1,1)$, and either  $(1,-1)$ or $(-1,-1)$.  

Let us discuss  how the contributions of $(\mu,\,\nu)=(\pm1,-1)$ sectors are computed.
Although the holonomies do not commute in $SU(2)$, they do commute in $SO(3)$, and so can be simultaneously diagonalized there.  We find, as was also mentioned in the previous section, that for some convenient choice of basis

\eqn\basishsutwo{ h = \pmatrix{-1 & 0 & 0 \cr 0 & 1 & 0 \cr 0 & 0 & -1}, \;\;\;\qquad
 g=\pmatrix{-1 & 0 & 0 \cr 0 & -1 & 0 \cr 0 & 0 & 1} \,.}
Thus, the lens space index for the ${\cal N}=1$ chiral field in the vector (or equivalently, adjoint) representation of $SO(3)$ with almost commuting holonomies can be written as
\eqn\almostCom{\eqalign{
&{\cal I}^{(R)}_{\chi,(ac)}(u)\equiv \left((p\,q)^{1-R}u^{-2}\right)^{\frac{r}{8}}\;
\times \cr
&\;\Gamma(\pm(p\,q)^{\frac{R}{2}}\,q^{\frac{r}2}u;q^r,\,pq)\,
\Gamma(\pm(p\,q)^{\frac{R}{2}}\,p^{\frac{r}{2}}u;p^r,\,pq)\,
\Gamma(-(p\,q)^{\frac{R}{2}}\,q^{r}u;q^r,\,pq)\,
\Gamma(-(p\,q)^{\frac{R}{2}}u;p^r,\,pq)\,.
}
} and the index of the vector multiplet of $SO(3)$ with the almost commuting holonomies is 
\eqn\almostComV{\eqalign{
&{\cal I}^{V}_{(ac)}\equiv \left(p\,q\right)^{-\frac{r}{8}}\;\times \cr
&\quad\frac1{\Gamma(\pm\,q^{\frac{r}2};q^r,\,pq)\,
\Gamma(\pm\,p^{\frac{r}{2}};p^r,\,pq)\,
\Gamma(-1;q^r,\,pq)\,
\Gamma(-p^r;p^r,\,pq)}\,.
}
}
There is no integral over the choice of $h$ to be taken here, unlike for sectors $(\pm1,1)$, since the set of flat connections is a single point.  Thus the contribution from  sector $(\pm1,-1)$ is simply a product of  contributions from all the chiral and vector fields of the theory.
 
Following the general discussion of the previous section, we can now write the lens indices of the three possible theories with the 
$su(2)$ Lie algebra \readinglines\ as
\eqn\sutwotheoriesB{
\eqalign{
r=0 \; {\rm mod}\; 4:\qquad& {\cal I}_{SU(2)} = Z_{1,1} \cr
& \,{\cal I}_{SO(3)_+} = \frac{1}{2}( Z_{1,1} + Z_{-1,1} + Z_{1,-1}+Z_{-1,-1} ) \cr
& \,{\cal I}_{SO(3)_-} = \frac{1}{2} ( Z_{1,1} + Z_{-1,1} + Z_{1,-1}-Z_{-1,-1} )\,.
}}

\eqn\sutwotheoriesA{
\eqalign{
r=2 \; {\rm mod}\; 4:\qquad & {\cal I}_{SU(2)} = Z_{1,1} \cr
& \,{\cal I}_{SO(3)_+} = \frac{1}{2} ( Z_{1,1} + Z_{-1,1} + Z_{-1,-1}+Z_{1,-1} ) \cr
& \, {\cal I}_{SO(3)_-} = \frac{1}{2} ( Z_{1,1} + Z_{-1,1} + Z_{-1,-1}-Z_{1,-1} )\,.
}}
Note that the $SU(2)$ version of the theory receives contributions only from the trivial $SO(3)$ bundle, {\it i.e.}, the one which lifts to an $SU(2)$ bundle.  
The theories with $SO(N)_\pm$ gauge groups differ by a discrete theta angle parameter, which weighs the contribution $Z_{\mu,\nu}$ for which $P_2(\mu,\nu)$ is non-zero with a factor $\pm 1$.  However, in the present case, $N=3$, these sectors are absent, and so the indices for the $SO(3)_\pm$ theories are equal.
 This follows from the fact that they are related by a shift of the theta angle by $\theta \rightarrow \theta+2 \pi$.  Since the index is not sensitive to the $\theta$ angle, it does not distinguish theories which are related in this way.  We will see a similar effect when we discuss  $SU(N)/\Z_N$ theories in section 5. However, there is a non-trivial difference between the $SU(2)$ calculation and the $SO(3)_\pm$ one, with the latter involving the almost commuting holonomies.  

If we have a duality which exchanges theories with different global structures, it will imply certain identities among the various indices above.  Let us now demonstrate this by performing explicit computations of the lens space index for $su(2)$ gauge theories, first for ${\cal N}=4$ SYM and then for ${\cal N}=1$ SQCD.

\subsec{${\cal N}=4 \; su(2)$ super Yang-Mills Theory}

As our first example of matching lens indices across dualities, consider ${\cal N}=4$ super Yang Mills (SYM) theory with Lie algebra $su(2)$.  Here the  theory contains a vector multiplet and $3$ chiral multiplets in the adjoint representation.  In the ${\cal N}=1$ notation, the latter have $R$-charge $2/3$.  Since the matter is in the adjoint representation, which is not acted on by the center, this is an example of a theory where we can allow non-trivial bundles.  Indeed this is true of ${\cal N}=4$ SYM for any gauge group, and we will study more examples in section~5.

As discussed in detail in \readinglines, there are several different versions of ${\cal N}=4$ SYM with a given Lie algebra. First one makes a choice of the gauge group, $SU(2)$ or $SO(3)$ in our case. 
Second, in the case of $SO(3)$ a discrete theta parameter can be turned:  value $0$ of which 
 corresponds to theory $SO(3)_+$ and value $1$ to $SO(3)_-$. These choices are correlated 
with a choice of allowed line operators in the given theory.
$S$-duality, which exchanges electric and magnetic charges,
can be seen to map the $SU(2)$ theory to the $SO(3)_+$ theory, while mapping the $SO(3)_-$ theory to itself.

Let us now turn to the lens index computation for these three theories.  The
 contributions of the four sectors, labeled by $(\mu,\nu)$, take the following explicit form
\eqn\sutwosectorsA{\eqalign{
(1,1):\qquad&{Z}_{1,1}=\sum_{m=0}^{\frac{r}2} \oint \frac{dz}{2\pi i z} \left(\frac12(1-z^{\pm2})\right)^{\delta_{[2m],0}}\,
{\cal I}_V([2m],z^2){\cal I}_V([-2m],z^{-2}){\cal I}_V(0,1)\times\cr
&\qquad\qquad\qquad\prod_{i=1}^3{\cal I}_\chi^{(\frac{2}{3})}([2m], u_i\, z^2)
{\cal I}_\chi^{(\frac23)}([-2m], u_i^{}\, z^{-2}){\cal I}_\chi^{(\frac23)}(0, u_i^{})\,,\cr
(-1,1):\qquad&{Z}_{-1,1}=\sum_{m=\frac12}^{\frac{r-1}2} \oint \frac{dz}{2\pi i z}\,
{\cal I}_V(m,z^2){\cal I}_V([-m],z^{-2}){\cal I}_V(0,1)\times\cr
&\qquad\qquad\qquad\prod_{i=1}^3{\cal I}_\chi^{(\frac{2}{3})}(m, u_i\, z^2)
{\cal I}_\chi^{(\frac23)}([-m], u_i^{}\, z^{-2})
{\cal I}_\chi^{(\frac23)}(0, u_i^{})\,,\cr
(-1,-1):\quad &{Z}_{-1,-1}=
{\cal I}^V_{(ac)}\prod_{i=1}^3{\cal I}_{\chi,(ac)}^{(\frac{2}{3})}(u_i)\,,\quad (1,-1)\;:\quad { Z}_{1,-1}=0\,,\quad
(\;r=\,2\;{\rm mod}\;4\;)\,,\cr
(1,-1):\quad &{Z}_{1,-1}=
{\cal I}^V_{(ac)}\prod_{i=1}^3{\cal I}_{\chi,(ac)}^{(\frac{2}{3})}(u_i)\,,\quad (-1,-1)\;:\quad { Z}_{-1,-1}=0\,,\quad
(\;r=\,0\;{\rm mod}\;4\;)\,.
}}
Here $u_i$ are fugacities coupling to a combination of the ${\cal N}=4$ R-symmetry.
We defined also
\eqn\defmod{
[m] \equiv n\; |\;0\leq n\leq r-1,\; n=m\;({\rm mod} \; r)\,.
}
Let us give the result of the computation for a few values of $r$. For $r=2$ we get
for the different $Z_{\mu,\nu}$,

\eqn\Aham{
\vbox{\offinterlineskip\tabskip=0pt
\halign{\strut\vrule#
&~$#$~\hfil\vrule
&~$#$~\hfil\vrule
&~$#$~\hfil\vrule
&~$#$~\hfil
&~$#$\hfil
&~$#$\hfil
&\vrule#
\cr
\noalign{\hrule}
&  \nu \backslash \mu&\qquad\qquad 1 & \qquad\qquad -1\cr
\noalign{\hrule}
&  \;\;\;1
  &   2+12 x^{4/3}-16 x^2+ & 1+3 x^{2/3}+3 x^{4/3}+3 x^2-\cr
&         &   \;\;\;\;+18 x^{8/3}-6 x^{10/3}+4 x^4+\dots & \;\;\;\;-9 x^{8/3}+12 x^{10/3}+38 x^4+\dots\cr
\noalign{\hrule}
& -1     &    & 1-3 x^{2/3}+9 x^{4/3}-19 x^2+\cr
&       &   \;\;\;\;\;\;\;\;\;0  &\;\;\;\;+27 x^{8/3}-18 x^{10/3}-34 x^4 +\dots\cr
}\hrule}}
Our results hold for any value of the fugacities but here we have refined the index only with $p=q=x$ for brevity. It is easy to see that the following equality holds

\eqn\eqSdual{
{Z}_{1,1}= \frac{1}{2}({Z}_{1,1}+{Z}_{-1,1}+{Z}_{-1,-1}+ {Z}_{1,-1} )\,.
} On the left-hand-side we have the lens space index of the theory with gauge group $SU(2)$ and
on the right-hand-side we have the theory with gauge group $SO(3)_+$.
In this case the index of $SO(3)_+$ is the same as the one for $SO(3)_-$, as
expected, since $Z_{1,-1}=0$.
 This equality is a confirmation
of ${\cal N}=4$ S-duality.\foot{Attached to this paper is a {\it Mathematica} notebook using which one can generate the results for any values of $r$ and with a maximal set of fugacities turned on.}

Next let us give also the result for $r=4$.
\eqn\AhamB{
\vbox{\offinterlineskip\tabskip=0pt
\halign{\strut\vrule#
&~$#$~\hfil\vrule
&~$#$~\hfil\vrule
&~$#$~\hfil\vrule
&~$#$~\hfil
&~$#$\hfil
&~$#$\hfil
&\vrule#
\cr
\noalign{\hrule}
&   \nu \backslash \mu& \qquad\qquad1 &\qquad\qquad-1\cr
\noalign{\hrule}
&  \;\;\;1
  &   3+3 x^{2/3}+15 x^{4/3}-15 x^2+ & 2+6 x^{2/3}+6 x^{4/3}+\cr
&         &   \;\;\;\;+33 x^{8/3}-36 x^{10/3}+76 x^4+\dots &\;\;\;\;+4 x^2+12 x^{10/3}+20 x^4+\dots\cr
\noalign{\hrule}
& -1         &1-3 x^{2/3}+9 x^{4/3}-19 x^2+&\cr
&        & \;\;\;\;+33 x^{8/3}-48 x^{10/3}+56 x^4+\dots&   \;\;\;\;\;\;\;\;\;0 \cr
}\hrule}}
This is also consistent with S-duality since
as one can check that the following holds 

\eqn\eqSdualB{
{Z}_{1,1}=\frac{1}{2} ({Z}_{1,1}+{Z}_{-1,1}+{Z}_{1,-1}+ {Z}_{-1,-1})\,.
} 

\

\subsec{${\cal N}=1 \; so(3)$ Seiberg Duality}

Let us now discuss the Seiberg duality of $Spin(3)$ ${\cal N}=1$ SQCD with $N_f=2$, {\it i.e.},
with two chiral fields in the vector representation of $Spin(3)$. This theory 
is dual to $SO(3)_-$ SQCD with $N_f=2$ with additional singlet mesonic fields with a superpotential~\SeibergPQ.
The charges of the singlet mesons here are consistent with a mass term and thus they do not contribute 
to the computation of the lens space index. Thus, as opposed to the ${\cal N}=4$ model discussed above we have one less chiral field and different R-charges. At the level of the lens space index the current 
case can be obtained from the former one by tuning the fugacities $u_i$ in an appropriate way: {\it e.g.} $u_3=x^{\frac13}$, $u_1=x^{-\frac1{6}}\,a$, $u_2=x^{-\frac1{6}}\,a^{-1}$.
For the $r=2$ case one obtains 
\eqn\SeibergBasic{\eqalign{
&
\vbox{\offinterlineskip\tabskip=0pt
\halign{\strut\vrule#
&~$#$~\hfil\vrule
&~$#$~\hfil\vrule
&~$#$~\hfil\vrule
&~$#$~\hfil
&~$#$\hfil
&~$#$\hfil
&\vrule#
\cr
\noalign{\hrule}
&  \nu \backslash \mu& \qquad\qquad1 & \qquad\qquad-1\cr
\noalign{\hrule}
&  \;\;\;1
  &  2+6 x+4 x^2-12 x^{5/2}+ &1+2 \sqrt{x}+3 x+2 x^{3/2}+3 x^2-\cr
&         &   \;\;\;\;+18 x^3-4 x^{7/2}+30 x^4+\dots & \;\;\;\;-6 x^{5/2}+2 x^3+16 x^{7/2}+18 x^4+\dots\cr
\noalign{\hrule}
& -1       &    & 1-2 \sqrt{x}+3 x-2 x^{3/2}+x^2- \cr
&       &   \;\;\;\;\;\;\;\;\;0  &\;\;\;\;-6 x^{5/2}+16 x^3-20 x^{7/2}+12 x^4+\dots\cr
}\hrule}
}
}
Consistently with Seiberg duality we find
\eqn\eqSeibdual{
{Z}_{1,1}=\frac{1}{2}({Z}_{1,1}+{Z}_{-1,1}+{Z}_{-1,-1}- {Z}_{1,-1})\,.
} The left-hand side is the index of the theory with $SU(2)$ gauge group and the right-hand side
is the index of the theory with $SO(3)_-$ gauge group (which again is the same as $SO(3)_+$ here).

Equipped with the technology of how to compute lens space index for theories with a non simply connected 
group we now turn to state our main results.

\

\newsec{$so(N_c)$ Seiberg duality}

We are now ready to discuss more intricate dualities recently suggested in~\readinglines.
We start with  Seiberg duality relating ${\cal N}=1$ gauge theories with $so(N_c)$ Lie algebra
and vector flavors~\refs{\SeibergPQ,\IntriligatorID}.

On side A of the duality we have $so(N_c)$ SQCD with $N_f$ flavors, $Q_i$,
in the vector representation, and no superpotential. The R-charge of the quark flavors $Q_i$ is 
$R_Q=\frac{N_f-N_c+2}{N_f}$. On side B of the duality we have $so(N_f-N_c+4)$ SQCD
with $N_f$ vector flavors, $q_i$, $\frac{N_f(N_f+1)}{2}$ gauge singlets $M_{ij}$, and a superpotential
$W=Mqq$. The R-charges of the dual quarks are $R_q=1-R_Q$, and the singlets $M_{ij}$
have R-charge $R_M=2\,R_Q$.
It has been argued in~\readinglines\ that
 there are actually {\it three} different dualities with this matter content
and Lie algebra differing by the global structure of the gauge group and 
discrete theta angle parameters,

\eqn\ASTSO{\eqalign{
&Spin(N_c) \qquad \longleftrightarrow  \qquad SO(N_f-N_c+4)_{-}\,,\cr
&SO(N_c)_{-} \qquad \longleftrightarrow  \qquad Spin(N_f-N_c+4)\,,\cr
&SO(N_c)_{+} \qquad \longleftrightarrow  \qquad SO(N_f-N_c+4)_{+}\,.
}
}
The lens indices of the dual pairs should match.  As in the previous section, to be able to see the global structure of the gauge group we must work on $L(r,1) \times \S^1$ with even $r$.\foot{For odd $r$, it is still a non-trivial identity that the indices match, but since we are mostly interested in the global structure of the gauge group we do not discuss these cases here.}  Then, as discussed after \pontsq, the lens indices of the different theories with $so(N_c)$ Lie algebra are given by

\eqn\soNtheories{
\eqalign{
& {\cal I}_{Spin(N_c)} = {Z}_{1,1}\,, \cr
& \;{\cal I}_{SO(N_c)_+}  = \frac{1}{2}( { Z}_{1,1} + {Z}_{-1,1} + { Z}_{1,-1}+
{Z}_{-1,-1})\,, \cr
& \;{\cal I}_{SO(N_c)_-} = \frac{1}{2} ( {Z}_{1,1} +{ Z}_{-1,1} + {Z}_{-1,-1}-{Z}_{1,-1} ),\qquad 
r=\,2\;{\rm mod}\;4\,,\cr 
& \;{\cal I}_{SO(N_c)_-} = \frac{1}{2} ( {  Z}_{1,1}+ {Z}_{-1,1} + { Z}_{1,-1}-{Z}_{-1,-1} ),\qquad 
r=\,0\;{\rm mod}\;4\,.
}}  Using~\pontsq, depending on the value of $r\,({\rm mod}\,4)$, one of the sectors $Z_{\pm1,-1}$
has a non-trivial Pontryagin square and thus is weighed with a minus sign when computing the $SO(N_c)_-$
partition function. Unlike the $so(3)$ case discussed in the previous section, for $so(N_c>4)$ all  four $Z_{\mu,\nu}$ are in general non-vanishing, thus  allowing us to distinguish $SO(N_c)_+$,
$SO(N_c)_-$, and $Spin(N_c)$.

It is useful to define the following quantities
\eqn\ems{
\eqalign{
&Y_{0,0}=\frac{Z_{1,1}+Z_{1,-1}}{2}\,,\qquad Y_{1,0}=\frac{Z_{1,1}-Z_{1,-1}}{2}\,,\cr
&Y_{0,1}=\frac{Z_{-1,1}+Z_{-1,-1}}{2}\,,\qquad Y_{1,1}=\frac{Z_{-1,1}-Z_{-1,-1}}{2}\,.
}
}
The dualities~\ASTSO\ imply thus the following identities. First for $r=2\;{\rm mod}\;4$

\eqn\idenSeib{
\eqalign{
&Y_{0,0}+Y_{1,0}={\widetilde Y}_{1,0}+{\widetilde Y}_{0,1},\cr
&Y_{1,0}+Y_{0,1}={\widetilde Y}_{0,0}+{\widetilde Y}_{1,0},\cr
&Y_{0,0}+Y_{0,1}={\widetilde Y}_{0,0}+{\widetilde Y}_{0,1}.
}
} The variables with the {\it tilde} are for the dual theory. 
There is a simple solution to this equations
\eqn\simpleY{
Y_{1,0}={\widetilde Y}_{1,0}\,,\qquad 
Y_{0,0}={\widetilde Y}_{0,1}\,,\qquad
Y_{0,1}={\widetilde Y}_{0,0}\,,\qquad Y_{1,1}={\widetilde Y}_{1,1}\,.
} The last equality does not follow from \idenSeib\ but
in fact we  find  that this is exactly the solution realized in practice. 
We will  comment on this fact in the summary section. For $r=0\;{\rm mod}\;4$ we similarly obtain

\eqn\idenSeibB{
\eqalign{
&Y_{0,0}+Y_{1,0}={\widetilde Y}_{0,0}+{\widetilde Y}_{1,1},\cr
&Y_{0,0}+Y_{1,1}={\widetilde Y}_{0,0}+{\widetilde Y}_{1,0},\cr
&Y_{0,0}+Y_{0,1}={\widetilde Y}_{0,0}+{\widetilde Y}_{0,1}.
}
} The solution to these equations which is realized in practice is
\eqn\simpleYB{
Y_{0,0}={\widetilde Y}_{0,0}\,,\qquad 
Y_{0,1}={\widetilde Y}_{0,1}\,,\qquad
Y_{1,1}={\widetilde Y}_{1,0}\,,\qquad
Y_{1,0}={\widetilde Y}_{1,1}\,.
} Note that this is different from~\simpleY.
From now on we will quote the results for, and implicitly assume that, $r=2$ unless otherwise explicitly stated.
The fact that the above identities hold is a highly non trivial check of the dualities~\ASTSO.

\noindent Let us now write explicit expressions for the  ${Z}_{\mu,\nu}$s and then use those
to check \simpleY. As above, we will quote the results for the index with a single fugacity $x$
for the sake of brevity. All our results hold with maximal set of fugacities turned on. An interested reader can generate these using the supplemented {\it Mathematica} notebook.

\

\noindent {\it The computation of ${Z}_{1,1}$}

\

By definition $Z_{1,1}$ is defined as a sum over $Spin(N_c)$ holonomies, {\it i.e.} $\mu=\nu=1$.  Let us write $N_c=2n+\epsilon$, where $n$ is the rank of the group, and $\epsilon=0$ or $1$.
Then for the $so(2n+\epsilon)$ theory this sector is given by:

\eqn\ZaSO{\eqalign{
&{Z}_{1,1}=\left(\prod_{a\leq b=1}^{N_f}
{\cal I}_\chi^{(R_M)}(0, u_a\,u_b)
\right)^{\epsilon_B}\times\cr
&\sum_{m_1,\cdots,m_n=0|\; \sum_{i=1}^n m_i=0\,({\rm mod}\,2)}^{\frac{r}2}
\, \hat n_{\bf m}\; {\cal I_V}(0,1)^n \oint \prod_{i=1}^n\frac{dz_i}{2\pi i z_i} 
\, \Delta^{(n,\epsilon)}_{{\mb m}}({\mb z})\,
\left(\prod_{i=1}^n{\cal I}_V([\pm m_i],(z_i)^{\pm1})\right)^\epsilon\times\cr
&\cr
&\qquad\prod_{i< j}\left({\cal I}_V([\pm(m_i-m_j)],(z_i/z_j)^{\pm 1})
{\cal I}_V([\pm(m_i+m_j)],(z_i\,z_j)^{\pm 1})\right)\times\cr
&\qquad\qquad\qquad\left(\prod_{i=1}^n\prod_{a=1}^{N_f}
{\cal I}_\chi^{(R)}([\pm m_i], u_a z_i)\right)\,
\left(\prod_{a=1}^{N_f}
{\cal I}_\chi^{(R)}(0, u_a)\right)^{\epsilon}\,.
}
} 
As before, we introduced fugacities, $u_a$,
for the $SU(N_f)$ flavor symmetry. Here $\epsilon_B$ is $0$ if we are on side A of the duality and $B$ if we are on side B, {\it i.e.}
it indicates whether singlet mesonic fields are included.
We are defining the computation here in the $SO(N_c)$ language, 
and in particular the holonomies here are those of $SO(N_c)$ vector representation. 
To make contact to our general definition of $Z_{\mu,\nu}$~\zmunudef\ the factor of $\hat n_{\bf m}$ was introduced: it is equal to $1$ if the $SO$ holonomy $g$ in vector representation has two lifts to $Spin$
which are related by Weyl symmetry, and equals $2$ if the two lifts are Weyl inequivalent.\foot{
A simple example of this factor can be seen by considering $su(2)$ duality of the previous section in $so(3)$ language. In $su(2)$ language the commuting holonomies when $r=2$ are ${\bf m}=(0,0),\,(1,1),\,(\frac12,\frac32)$ (in the notations of the previous section these are $m=0,1,\frac12$), whereas in $so(3)$ language these are ${\bf m}'=(0,0,0),\,(1,0,1)$.
Here, $(0,0,0)$ lifts to $(0,0)$ and $(1,1)$ in the $su(2)$ language, but $(1,0,1)$ lifts to two elements 
related by Weyl symmetry, $(\frac12,\frac32)$ and $(\frac32,\frac12)$. 
} 
To fix the Weyl symmetry of the $m_i$ we always assume that
\eqn\mis{\eqalign{
&\epsilon=1\;:\qquad\frac{r}{2}\geq m_1\geq m_2\geq \cdots\geq m_n\geq 0\,,\cr
&\epsilon=0\;:\qquad\frac{r}{2}\geq m_1\geq m_2\geq \cdots\geq ||m_n||\geq 0\,.
}
} Here $||m||$ equals $m$ if $0\leq m\leq r/2$ and equals $r-m$ if  $r/2\leq m<r$.
 The R-charge $R$ is either $R_Q$ or $R_q$
depending on the side of the duality. The Haar measure
$\Delta^{(n,\epsilon)}_{{\mb m}}({\mb z})$ is defined as
\eqn\HaarSO{\eqalign{
&\Delta^{(n,\epsilon)}_{{\mb m}}({\mb z})=
\frac1{|W_{{\mb m},n,\epsilon}|}\,
\prod_{i< j}^n\left((1-z_iz_j)(1-\frac{1}{z_iz_j})\right)^{\delta_{[m_i+m_j],0}}
\left((1-\frac{z_i}{z_j})(1-\frac{z_j}{z_i})\right)^{\delta_{[m_i-m_j],0}}\times\cr
&\qquad\qquad \left(\prod_{j=1}^n\left((1-z_j)(1-\frac1{z_j})\right)^{\delta_{[m_j],0}}\right)^\epsilon\,,
}
} where $|W_{{\mb m},n,\epsilon}|$ is the size of the unbroken Weyl group determined by
demanding
\eqn\WeylDet{
\oint \prod_{i=1}^n\frac{dz_i}{2\pi i z_i}\,\Delta^{(n,\epsilon)}_{{\mb m}}({\mb z})=1\,.
}
\

\noindent {\it The computation of ${Z}_{-1,1}$}

\

Here we sum over holonomies which  have $\mu=-1$ and $\nu=1$.
For  $so(2n+\epsilon)$ theory ${Z}_{-1,1}$ is given by
\eqn\ZbSO{\eqalign{
&{Z}_{-1,1}=\left(\prod_{a\leq b=1}^{N_f}
{\cal I}_\chi^{(R_M)}(0, u_a\,u_b)
\right)^{\epsilon_B}\times\cr
&\sum_{m_1,\cdots,m_n=0|\, \sum_{i=1}^n m_i=1\,({\rm mod}\,2)}^{r/2}
\,\hat n_{\bf m}\; {\cal I_V}(0,1)^n \oint \prod_{i=1}^n\frac{dz_i}{2\pi i z_i} 
\, \Delta^{(n,\epsilon)}_{{\mb m}}({\mb z})\,
\left(\prod_{i=1}^n{\cal I}_V([\pm m_i],(z_i)^{\pm1})\right)^\epsilon\times\cr
&\cr
&\qquad\prod_{i< j}\left({\cal I}_V([\pm(m_i-m_j)],(z_i/z_j)^{\pm1})
{\cal I}_V([\pm(m_i+m_j)],(z_i\,z_j)^{\pm1})\right)\times\cr
&\qquad\qquad\qquad\left(\prod_{i=1}^n\prod_{a=1}^{N_f}
{\cal I}_\chi^{(R)}([\pm m_i], u_a\, z_i)\right)\,
\left(\prod_{a=1}^{N_f}
{\cal I}_\chi^{(R)}(0, u_a)\right)^{\epsilon}\,.
}
}

\

\noindent {\it The computation of ${Z}_{\pm1,-1}$}

\

Here we have to sum over holonomies with $\nu=-1$.
A procedure of doing so for arbitrary groups was discussed in~\refs{\SchweigertTG,\almost}.
We claim that the answer for the $so(N_c)$ case is given by a simple generalization of  the procedure for doing so we discussed in the previous section for $SO(3)$: 
one obtains that the holonomies $h$ and $g$ satisfying $\nu=-1$ can be always chosen  (by
gauge transformations) to be
\eqn\almost{
 g=\pmatrix{-1 & 0 & 0 &0\cr 0 & -1 & 0 &0\cr 0 & 0 & 1&0\cr 0&0&0&{\cal G}} , \;\;\;\qquad
h = \pmatrix{-1 & 0 & 0 & 0 \cr 0 & 1 & 0&0 \cr 0 & 0 & -1&0\cr 0&0&0&{\cal H}}\,.
}
Here ${\cal H}$ and ${\cal G}$ are elements of $SO(N_c-3)$ 
such that they commute in $Spin(N_c-3)$. Further the distinction between $\mu=1$ and
$\mu=-1$ is whether  $g^r=1$ lifts to $1$ or $-1$ in $Spin(N_c)$:
this means that ${\cal G}^r$ lifts to $1$ ($-1$) or $-1$ ($1$)  in $Spin(N_c-3)$
respectively in case that $r=2\, ({\rm mod}\, 4)$ ($r=0\, ({\rm mod}\, 4)$).  

\noindent We thus can write ${Z}_{1,-1}$ as follows
\eqn\ZcSO{\eqalign{
&{Z}_{1,-1}=\left(\prod_{a\leq b=1}^{N_f}
{\cal I}_\chi^{(R_M)}(0, u_a\,u_b)
\right)^{\epsilon_B}\,\left(\prod_{a=1}^{N_f}
{\cal I}_\chi^{(R)}(0, u_a)\right)^{\epsilon}\times\cr
&\sum_{m_1,\cdots,m_{n-2+\epsilon}=0|\, \sum_{i=1}^{n-2+\epsilon} m_i=0\,({\rm mod}\,2)}^{r/2} {\cal I_V}(0,1)^{n-2+\epsilon} \oint \prod_{i=1}^{n-2+\epsilon} \frac{dz_i}{2\pi i z_i} 
\, \widetilde \Delta^{(n,\epsilon)}_{{\mb m}}({\mb z})\,
\times\cr
&\qquad \left(\prod_{i=1}^{n-2+\epsilon}{\cal I}_V([\pm m_i],(z_i)^{\pm1})\right)^{1-\epsilon}\;\left({\cal I}_V(\frac{r}2, 1)
{\cal I}_V(\frac{r}2, -1)
{\cal I}_V(0, -1)\right)^{1+\epsilon}\times\cr
&\qquad\prod_{i< j}^{n-2+\epsilon}\left({\cal I}_V([\pm(m_i-m_j)],(z_i/z_j)^{\pm1})
{\cal I}_V([\pm(m_i+m_j)],(z_i\,z_j)^{\pm1})\right)\times\cr
&\qquad\prod_{j=1}^{n-2+\epsilon}\left(
{\cal I}_V([\frac{r}2\pm m_j],(z_j)^{\pm1})
{\cal I}_V([\pm m_j],(-z_j)^{\pm1})
{\cal I}_V([\frac{r}2\pm m_j],(-z_j)^{\pm1})\right)\times
\cr
&\qquad\prod_{a=1}^{N_f}\left(
{\cal I}_\chi^{(R)}(\frac{r}2, u_a)
{\cal I}_\chi^{(R)}(\frac{r}2, -u_a)
{\cal I}_\chi^{(R)}(0, -u_a)
\prod_{i=1}^{n-2+\epsilon}
{\cal I}_\chi^{(R)}([\pm m_i], u_a\, z_i)\right)\,
\,.
}
} 
Here $\widetilde \Delta$ is given by
\eqn\haarspec{
\widetilde \Delta^{(n,\epsilon)}_{{\mb m}}({\mb z})=\frac{1}{\delta W_{{\mb m},n,\epsilon}}\,
\Delta^{(n-2+\epsilon,1-\epsilon)}_{{\mb m}}({\mb z})\,
\prod_{i=1}^{n-2+\epsilon}\left(1+z_j^{\pm1}\right)^{\delta_{[m_j],0}}
\left(1-z_j^{\pm2}\right)^{\delta_{[m_j+\frac{r}2],0}}\,,
} with $\delta W_{{\mb m},n,\epsilon}$ defined by demanding
\eqn\WeyltildeDet{
\oint \prod_{i=1}^{n-2+\epsilon}\frac{dz_i}{2\pi i z_i}\,\widetilde\Delta^{(n,\epsilon)}_{{\mb m}}({\mb z})=1\,.
}

 In the same way  ${Z}_{-1,-1}$ is given by
\eqn\ZdSO{\eqalign{
&{Z}_{-1,-1}=\left(\prod_{a\leq b=1}^{N_f}
{\cal I}_\chi^{(R_M)}(0, u_a\,u_b)
\right)^{\epsilon_B}\,\left(\prod_{a=1}^{N_f}
{\cal I}_\chi^{(R)}(0, u_a)\right)^{\epsilon}\times\cr
&\sum_{m_1,\cdots,m_{n-2+\epsilon}=0|\, \sum_{i=1}^{n-2+\epsilon} m_i=1\,({\rm mod}\,2)}^{r/2} {\cal I_V}(0,1)^{n-2+\epsilon}\oint \prod_{i=1}^{n-2+\epsilon} \frac{dz_i}{2\pi i z_i} 
\, \widetilde \Delta^{(n,\epsilon)}_{{\mb m}}({\mb z})\, \times\cr
&\qquad\left(\prod_{i=1}^{n-2+\epsilon}{\cal I}_V([\pm m_i],(z_i)^{\pm1})\right)^{1-\epsilon} \left({\cal I}_V(\frac{r}2, 1)
{\cal I}_V(\frac{r}2, -1)
{\cal I}_V(0, -1)\right)^{1+\epsilon}\times\cr
&\qquad\prod_{i< j}^{n-2+\epsilon}\left({\cal I}_V([\pm(m_i-m_j)],(z_i/z_j)^{\pm1})
{\cal I}_V([\pm(m_i+m_j)],(z_i\,z_j)^{\pm1})\right)\times\cr
&\qquad\prod_{j=1}^{n-2+\epsilon}\left(
{\cal I}_V([\frac{r}2\pm m_j],(z_j)^{\pm1})
{\cal I}_V([\pm m_j],(-z_j)^{\pm1})
{\cal I}_V([\frac{r}2\pm m_j],(-z_j)^{\pm1})\right)\times
\cr
&\qquad\prod_{a=1}^{N_f}\left(
{\cal I}_\chi^{(R)}(\frac{r}2, u_a)
{\cal I}_\chi^{(R)}(\frac{r}2, -u_a)
{\cal I}_\chi^{(R)}(0, -u_a)
\prod_{i=1}^{n-2+\epsilon}
{\cal I}_\chi^{(R)}([\pm m_i], u_a\, z_i)\right)\,
\,.
}
}

\subsec{Examples}

Let us next present  several typical examples of the lens space index
of Seiberg dual pairs. In all the examples one can explicitly see
that the identities \simpleY\ hold consistently with having
the dualities \ASTSO.

\

\noindent {\it{Example I: $so(3)\;\longleftrightarrow so(N_c)$ dualities}} \

\

The lens space index does not distinguish $SO(3)_+$ and $SO(3)_-$ groups, as discussed in the
previous section. Therefore the dualities~\ASTSO\  imply that the lens  index of a 
theory with $N_f$ flavors and $Spin(N_f+1)$ gauge group is the same as the index of a theory 
with $SO(N_f+1)_+$ gauge group. A similar situation also occurs for dualities where on one side we 
have an $so(4)\sim su(2)\times su(2)$ theory. There again the $SO_+$ and $SO_-$ theories are related
by a shift of the theta angle which the lens space index does not depend on. Then we expect that the lens indices of theory with $N_f$ flavors and $Spin(N_f)$ gauge group are equal to the one with $N_f$ 
flavors and $SO(N_f)_+$ gauge group.

Let us give an explicit example of $so(3)\leftrightarrow so(6)$ dualities.
The different $so(3)$ $N_f=5$ (without mesons) sectors $Z_{\mu,\nu}$ contribute as follows
\eqn\SOsixseven{\eqalign{
&
\vbox{\offinterlineskip\tabskip=0pt
\halign{\strut\vrule#
&~$#$~\hfil\vrule
&~$#$~\hfil\vrule
&~$#$~\hfil\vrule
&~$#$~\hfil
&~$#$\hfil
&~$#$\hfil
&\vrule#
\cr
\noalign{\hrule}
& \nu \backslash \mu  & \qquad\qquad 1 & \qquad\qquad-1\cr
\noalign{\hrule}
&  \;\;\;1      
  & 2+30 x^{8/5}-50 x^2+& 
1+5 x^{4/5}-5 x^{6/5}+15 x^{8/5}-\cr
&         &  +40 x^{12/5}-120 x^{14/5}+\dots
 & -23 x^2+45 x^{12/5}-90 x^{14/5}+\dots\cr
\noalign{\hrule}
& -1       & 
 0
   & 1-5 x^{4/5}+5 x^{6/5}+15 x^{8/5}-\cr
&       &  &-25 x^2-25 x^{12/5}+60 x^{14/5}+\dots\cr
}\hrule}
}
}
\
The different $so(6)$ $N_f=5$ (with mesons) sectors $\widetilde Z_{\mu,\nu}$ contribute as follows
\eqn\SOsixseven{\eqalign{
&
\vbox{\offinterlineskip\tabskip=0pt
\halign{\strut\vrule#
&~$#$~\hfil\vrule
&~$#$~\hfil\vrule
&~$#$~\hfil\vrule
&~$#$~\hfil
&~$#$\hfil
&~$#$\hfil
&\vrule#
\cr
\noalign{\hrule}
&  \nu\backslash\mu& \qquad\qquad1 & \qquad\qquad-1\cr
\noalign{\hrule}
&   \;\;\;1            
  & 2+30 x^{8/5}-49 x^2+& 
1+5 x^{4/5}-5 x^{6/5}+15 x^{8/5}-\cr
&         &  +30 x^{12/5}-75 x^{14/5}+\dots
 & -24 x^2+55 x^{12/5}-135 x^{14/5}+\dots\cr
\noalign{\hrule}
& -1       & 
x^2-10 x^{12/5}+45 x^{14/5}+\cdots 
   &1-5 x^{4/5}+5 x^{6/5}+15 x^{8/5}-\cr
&       &  &-26 x^2-15 x^{12/5}+15 x^{14/5}+\dots\cr
}\hrule}
}
}
\ In particular the equality of the lens indices with $Spin(6)$ and  $SO(6)_+$ gauge groups
implies that $\widetilde {Z}_{1,1}=\widetilde {Z}_{1,-1}+\widetilde {Z}_{-1,1}+
\widetilde {Z}_{-1,-1}$ which can be easily verified
to hold for explicit values in the table above.

\

\noindent {\it{Example II: $so(6)\;\longleftrightarrow so(6)$ dualities}}\

\

Next we consider duals of $so(6)$ theories with eight flavors.
The dual here has the same Lie algebra.
The computation of the lens indices for side A and side B differs only if we turn on 
fugacities for the $SU(N_f)$ global symmetry:
 the mesonic operators on side B of the duality
have R-charge one and thus do not contribute to the index if one does not turn on fugacities for flavor symmetries, since then the symmetries with which the lens index is refined
are consistent with turning on a mass term for mesons.
 Not  refining with flavor symmetries there is only one identity that the lens
index has to satisfy, ${Z}_{1,1}-{Z}_{-1,1}={Z}_{-1,-1}-{Z}_{1,-1}$, which can be seen to hold in the 
explicit table below. 
The different $so(6)$ $N_f=8$ sectors contribute as follows
\eqn\SOsixeight{\eqalign{
&
\vbox{\offinterlineskip\tabskip=0pt
\halign{\strut\vrule#
&~$#$~\hfil\vrule
&~$#$~\hfil\vrule
&~$#$~\hfil\vrule
&~$#$~\hfil
&~$#$\hfil
&~$#$\hfil
&\vrule#
\cr
\noalign{\hrule}
&  \nu\backslash \mu& \qquad\qquad1 & \qquad\qquad-1\cr
\noalign{\hrule}
&   \;\;\;1      
  &  2+72 x+1205 x^2+12712 x^3+& 
1 + 36 x + 674 x^2 + 8076 x^3+\cr
&         &   \;\;\;\;+96468 x^4+571432 x^5+\dots & \;\;\;\; + 67125 x^4 + 418564 x^5+\dots\cr
\noalign{\hrule}
& -1&
x^2+8 x^3+36 x^4+248 x^5+\cdots 
   &1 + 36 x + 532 x^2 + 4644 x^3+\cr
&       &  &\;\;\;\;
 + 29379 x^4 + 153116 x^5+\dots\cr
}\hrule}
}
}
\ In the following we list several  examples of  results for less degenerate cases where all the three types
of dualities~\ASTSO\ can be distinguished by the lens space index computation. In all these 
examples the identities~\simpleY\ are easily verified. 

\

\noindent {\it{Example III: $so(5)\;\longleftrightarrow so(6)$ dualities}}\

\

The different $so(5)$ $N_f=7$ (no mesons) sectors contribute as follows
\eqn\SOsixseven{\eqalign{
&
\vbox{\offinterlineskip\tabskip=0pt
\halign{\strut\vrule#
&~$#$~\hfil\vrule
&~$#$~\hfil\vrule
&~$#$~\hfil\vrule
&~$#$~\hfil
&~$#$\hfil
&~$#$\hfil
&\vrule#
\cr
\noalign{\hrule}
&  \nu\backslash\mu& \qquad\qquad1 & \qquad\qquad-1\cr
\noalign{\hrule}
&  \;\;\;1      
  & 2+56 x^{8/7}-97 x^2+812 x^{16/7}+& 
 1+28 x^{8/7}+35 x^{12/7}-47 x^2+406 x^{16/7}-\cr
&         &  +7 x^{18/7}+84 x^{20/7}-2506 x^{22/7}+\dots
 & -133 x^{18/7}+777 x^{20/7}-1211 x^{22/7}+\dots\cr
\noalign{\hrule}
& -1&
 x^2-7 x^{18/7}+28 x^{22/7}+\cdots 
   & 1+28 x^{8/7}-35 x^{12/7}-49 x^2+406 x^{16/7}+\cr
&       &  &+147 x^{18/7}-735 x^{20/7}-1267 x^{22/7}+\dots\cr
}\hrule}
}
}
\
The different $so(6)$ $N_f=7$ (with mesons) sectors contribute as follows
\eqn\SOsixseven{\eqalign{
&
\vbox{\offinterlineskip\tabskip=0pt
\halign{\strut\vrule#
&~$#$~\hfil\vrule
&~$#$~\hfil\vrule
&~$#$~\hfil\vrule
&~$#$~\hfil
&~$#$\hfil
&~$#$\hfil
&\vrule#
\cr
\noalign{\hrule}
&  \nu\backslash\mu& \qquad\qquad1 & \qquad\qquad-1\cr
\noalign{\hrule}
&   \;\;\;1      
  &  2+56 x^{8/7}-97 x^2+812 x^{16/7}+& 
1+28 x^{8/7}+35 x^{12/7}-47 x^2+406 x^{16/7}-\cr
&         &  +14 x^{18/7}+63 x^{20/7}-2506 x^{22/7}+\dots
 & -140 x^{18/7}+798 x^{20/7}-1211 x^{22/7}+\dots\cr
\noalign{\hrule}
& -1       & 
x^2-21 x^{20/7}+28 x^{22/7}+\cdots 
   &1+28 x^{8/7}-35 x^{12/7}-49 x^2+406 x^{16/7}+\cr
&       &  &+140 x^{18/7}-714 x^{20/7}-1267 x^{22/7}+\dots\cr
}\hrule}
}
}
\

\

\noindent {\it{Example IV: $so(5)\;\longleftrightarrow so(7)$ dualities}}\

\

The different $so(7)$ $N_f=8$ (no mesons) sectors contribute as follows
\eqn\SOsixnine{\eqalign{
&
\vbox{\offinterlineskip\tabskip=0pt
\halign{\strut\vrule#
&~$#$~\hfil\vrule
&~$#$~\hfil\vrule
&~$#$~\hfil\vrule
&~$#$~\hfil
&~$#$\hfil
&~$#$\hfil
&\vrule#
\cr
\noalign{\hrule}
& \nu\backslash\mu & \qquad\qquad1 & \qquad\qquad-1\cr
\noalign{\hrule}
&   \;\;\;1      
  & 2+72 x^{\frac34}+1332 x^{\frac32}-127 x^2+& 
1 + 36 x^{\frac34} + 666 x^{\frac32} + 56 x^{\frac{15}8} - 62 x^2 +\cr
&         &  +16872 x^{\frac94}+16 x^{\frac{21}8}-4300 x^{\frac{11}4}+\dots 
&+ 8436 x^{\frac94} + 
 1800 x^{\frac{21}8} - 2096 x^{\frac{11}4}+\dots\cr
\noalign{\hrule}
& -1&
x^2+36 x^{\frac{11}4}+\cdots 
   &1+36 x^{\frac34}+666 x^{\frac32}-56 x^{\frac{15}8}-64 x^2+\cr
&       &  &
+8436 x^{\frac94}-1800 x^{\frac{21}8}-2168 x^{\frac{11}4}+\dots\cr
}\hrule}
}
}
\
The different $so(5)$ $N_f=8$ (with mesons) sectors contribute as follows
\eqn\SOsixnine{\eqalign{
&
\vbox{\offinterlineskip\tabskip=0pt
\halign{\strut\vrule#
&~$#$~\hfil\vrule
&~$#$~\hfil\vrule
&~$#$~\hfil\vrule
&~$#$~\hfil
&~$#$\hfil
&~$#$\hfil
&\vrule#
\cr
\noalign{\hrule}
&  \nu\backslash\mu& \qquad\qquad1 & \qquad\qquad-1\cr
\noalign{\hrule}
&  \;\;\;1      
  &  2+72 x^{\frac34}+1332 x^{\frac32}-127 x^2+& 
1 + 36 x^{\frac34} + 666 x^{\frac32} + 56 x^{\frac{15}8} - 62 x^2 +\cr
&         &  +16872 x^{\frac94}+8 x^{\frac{21}8}-4300 x^{\frac{11}4}+\dots 
&+ 8436 x^{\frac94} + 
 1808 x^{\frac{21}8} - 2096 x^{\frac{11}4}+\dots\cr
\noalign{\hrule}
& -1&
x^2-8 x^{\frac{21}8}+36 x^{\frac{11}4}+\cdots 
   & 1+36 x^{\frac34}+666 x^{\frac32}-56 x^{\frac{15}8}-64 x^2+\cr
&       &  &
+8436 x^{\frac94}-1792 x^{\frac{21}8}-2168 x^{\frac{11}4}+\dots\cr
}\hrule}
}
}
\

\

\noindent {\it{Example V: $so(6)\;\longleftrightarrow so(7)$ dualities}}

\

The different $so(7)$ $N_f=9$ (no mesons) sectors contribute as follows
\eqn\SOsixnine{\eqalign{
&
\vbox{\offinterlineskip\tabskip=0pt
\halign{\strut\vrule#
&~$#$~\hfil\vrule
&~$#$~\hfil\vrule
&~$#$~\hfil\vrule
&~$#$~\hfil
&~$#$\hfil
&~$#$\hfil
&\vrule#
\cr
\noalign{\hrule}
&  \nu\backslash \mu& \qquad\qquad1 & \qquad\qquad-1\cr
\noalign{\hrule}
&   \;\;\;1      
  & 2+90 x^{\frac89}+2070 x^{\frac{16}9}-161 x^2+& 
1+45 x^{\frac89}+1035 x^{\frac{16}9}-79 x^2+\cr
&         &  +32430 x^{\frac83}-6903 x^{\frac{26}9}+144 x^{\frac{28}9}+\dots 
&+126 x^{\frac{20}9}+16215 x^{\frac83}-3384 x^{\frac{26}9}+4986 x^{\frac{28}9}+\dots\cr
\noalign{\hrule}
& -1&
x^2+45 x^{\frac{26}9}+\cdots 
   & 1+45 x^{\frac89}+1035 x^{\frac{16}9}-81 x^2-\cr
&       &  &
-126 x^{\frac{20}9}+16215 x^{\frac83}-3474 x^{\frac{26}9}-4914 x^{\frac{28}9}+\dots\cr
}\hrule}
}
}
\
The different $so(6)$ $N_f=9$ (with mesons) sectors contribute as follows
\eqn\SOsixnine{\eqalign{
&
\vbox{\offinterlineskip\tabskip=0pt
\halign{\strut\vrule#
&~$#$~\hfil\vrule
&~$#$~\hfil\vrule
&~$#$~\hfil\vrule
&~$#$~\hfil
&~$#$\hfil
&~$#$\hfil
&\vrule#
\cr
\noalign{\hrule}
&  \nu\backslash\mu& \qquad\qquad1 & \qquad\qquad-1\cr
\noalign{\hrule}
&  \;\;\;1      
  &  2+90 x^{\frac89}+2070 x^{\frac{16}9} -161 x^2& 
1+45 x^{\frac89}+1035 x^{\frac{16}9}-79 x^2+\cr
&         &   +32430 x^{\frac83}-6903 x^{\frac{26}9}+108 x^{\frac{28}9}+\dots &
+126 x^{\frac{20}9}+16215 x^{\frac83}-3384 x^{\frac{26}9}+5022 x^{\frac{28}9}+\dots\cr
\noalign{\hrule}
& -1&
x^2+45 x^{\frac{26}9}-36 x^{\frac{28}9}+\cdots 
   &1+45 x^{\frac89}+1035 x^{\frac{16}9}-81 x^2-\cr
&       &  &
-126 x^{\frac{20}9}+16215 x^{\frac83}-3474 x^{\frac{26}9}-4878 x^{\frac{28}9}+\dots\cr
}\hrule}
}
}
\

\

To summarize, as can be explicitly deduced from the tables above, the identities~\simpleY\ following from dualities~\ASTSO\ are satisfied in all quoted examples.

\newsec{$S$-duality of ${\cal N}=4$ SYM  with $su(N)$ Lie algebra}

In this section we study the lens space index of ${\cal N}=4$ gauge theories, and in particular we perform certain checks of $S$-duality.  This is a non-perturbative duality which exchanges the gauge group $G$ with its Langlands dual $G^L$.  In the non-simply laced case, in general, the Lie algebra of $G^L$ is different from that of $G$, and so the matching of the supersymmetric index already provides a non-trivial test of the duality \refs{\GaddeKB,\SpiridonovQV}.   However, in the simply laced case, the only difference between $G$ and $G^L$ is in their global structure, so we must go to the lens space index to find a non-trivial test of this duality.  

In this section we will restrict the discussion
to the case of gauge groups based on $su(N)$ Lie algebra since it is rich enough to illustrate the general features, but simple enough so that the discussion can be made accesible and explicit.   The matter here is in the adjoint representation of $su(N)$, and thus we can discuss a wide variety of theories differing by the global structure of the group and/or by choices of discrete theta parameters: or, in the language of 
~\readinglines, by the choice of line operators.
We considered the relatively simple case of $su(2)$ theories above, and here will see  that the  higher rank cases exhibit much more interesting  series of duality moves that exchange the various global structures described in~\readinglines.  As will be shown the computation of the lens space index gives the same result for different theories in the same  S-duality orbit, while giving, at least in some case, different results for different orbits.

\subsec{ Classifying holonomies for the $su(N)$ theories.}

Let us start the discussion by studying the different sectors contributing to the lens space index
in the case the Lie algebra is $su(N)$.  We will derive all the needed ingredients for the lens space index computation  in this sub-section and then in the next subsection admix these to the general prescription of the previous sections to show that the index of the ${\cal N}=4$ SYM gauge theories is quite non-trivially consistent with S-duality.

The simply connected group with Lie algebra $su(N)$ is $SU(N)$, and its center is $Z=\Z_N$.  Since we will be studying theories with matter which is not charged under the full center, we can concentrate without loss on the gauge group $G$ being the maximal quotient, $SU(N)/\Z_N$:
the holonomies contributing to any other $SU(N)/\Z_{d\neq N}$ can be determined using \gfromgo.  Thus, following \liftItilde, to compute $Z_{\mu,\nu}$ we would like to find all gauge-inequivalent pairs of $SU(N)$ elements $\widetilde g,\;\widetilde h$ satisfying
\eqn\suAC{
{\widetilde g}^r = \mu, \qquad {\widetilde g}\, {\widetilde h} \,{\widetilde g}^{-1} {\widetilde h}^{-1}= \nu\,, 
}
where $\mu \in Z/Z^r$ and $\nu \in {\rm ker}(r)$.  In the present case both of these groups are isomorphic to $\Z_s$, where $s=(N,r)$, the greatest common divisor of $N$ and $r$.  Let us define $\zeta_N=e^{2\pi i/N}$.  Then it is convenient to define integers $k$ and $\ell$ by

\eqn\klint{ \mu = {\zeta_N}^k, \qquad\qquad\nu = {\zeta_N}^\ell\,,}
where the identification on $\mu$ and the condition $\nu^r=1$ allow us to write:

\eqn\kliden{ k \sim k + s, \qquad\qquad \ell = \hat{\ell} \;\frac{N}{s}\,,}
so that we can take $k ,\hat{\ell} \in \{0,...,s-1\}$.  Let us see which choices of $k$ and $\hat{\ell}$ can actually occur.

First, if $\hat{\ell}=0$, then ${\widetilde g}$ and ${\widetilde h}$ can be simultaneously diagonalized, and  we can solve the first equation in $\suAC$ for any $k$ by writing, for integers $m_i$,

\eqn\rildek{ {\widetilde g} = {\rm diag}(e^{\frac{2 \pi i }{r}(m_1+\frac{k}{N})},\;\cdots\;,e^{\frac{2 \pi i }{r}(m_N+\frac{k}{N})})\,.}
Here we impose $\sum_i m_i + k = 0\;({\rm mod}\;r)$ so ${\widetilde g}$ has determinant one.  We can fix the residual gauge (Weyl) symmetry by imposing $r>m_1 \geq ... \geq m_N \geq 0$.
The contribution of a chiral multiplet in the adjoint representation from this holonomy can be written using the standard weight basis as

\eqn\chisdual{ {\cal I}_\chi^{(R)}(0,1)^{N-1} \prod_{i \neq j} {\cal I}_\chi^{(R)}
([m_i - m_j], z_i {z_j}^{-1}) \,,}
where the prefactor comes from the contribution of the Cartan.  There is a similar contribution for the vector multiplet.  In addition, the Haar measure is given by

\eqn\suhaar{
\Delta_{{\bf m}}(z) = \frac{1}{|W_{{\bf m},N}|}\,\prod_{i<j} ((1- z_i {z_j}^{-1})(1- {z_i}^{-1} z_j))^{\delta_{[m_i - m_j],0}} \,.
} Here $|W_{{\bf m},N}|$ is the size of the un-broken Weyl group determined by demanding that the measure integrates to one.
Note that the dependence on $\mu={\zeta_N}^k$ enters the computation of the index through the constraint $\sum_i m_i+k=0 \; ({\rm mod} \; r)$.  Shifting $k$ by $N$ or $r$ does not affect this constraint, so the parameter $k$ is really defined modulo $s=(N,r)$.

Now let us consider solutions to \suAC\ for non-trivial $\nu$, {\it i.e.}, almost commuting holonomies.  To do this, it will be useful to introduce the so-called ``clock'' and ``shift'' matrices,

\eqn\csdef{ C_{\ell,N} = {\epsilon_N}^\ell \pmatrix{ 
1 & 0 & 0 & \cdots & 0 \cr
0 & {\zeta_N}^\ell & 0 & \cdots & 0 \cr
0 & 0 & {\zeta_N}^{2\ell} & \cdots & 0 \cr
\vdots & & &  & \vdots \cr
0 & 0 & 0 & \cdots &  {\zeta_N}^{(N-1)\ell}  }, \;\;\;
S_N = \epsilon_N \pmatrix{ 
0 & 1 & 0 & \cdots & 0 \cr
0 & 0 & 1 & \cdots & 0 \cr
0 & 0 & 0 & \cdots & 0 \cr
\vdots & & &  & \vdots \cr
1 & 0 & 0 & \cdots & 0}\,. }
Here $\epsilon_N$ is picked to ensure these have unit determinant.  We will take:

\eqn\epsilonN{
\epsilon_N =\cases{
1 & $N \; {\rm odd}$ \cr
e^{\pi i/N} & $N \; {\rm even}$ \cr
}
}
These are relevant because they satisfy:
\eqn\cseq{ C_{\ell,N}\; S_N = {\zeta_N}^\ell\; S_N\; C_{\ell,N}\,. }
If $(N,\ell)=1$, then these are the unique $U(N)$ matrices, up to simultaneous conjugation and scalar multiplication, which satisfy $A\, B = {\zeta_N}^\ell \,B\,A$.  This can be proved by an argument analogous to that used in section $3$ for the case $N=2$.  More generally, if we write $(N,\ell)=d$, we find the solutions to
\eqn\UNAC{
A\,B\, A^{-1} B^{-1} = {\zeta_N}^\ell\,,
}
can be written as:
\eqn\UNACsol{
A=  {A}_d\, \otimes \,C_{\ell/d,N/d}, \qquad\qquad B=  {B}_d\, \otimes\, S_{N/d}\,,
}
where $A_d$ and $B_d$ are arbitrary commuting matrices in $U(d)$, which can be taken 
to both be diagonal.  
Thus it is convenient to organize the solutions to \suAC\ by the value of $d \equiv (N,\ell)$.

Let us first consider the case $d=1$.  This can only occur if $N/s=1$, {\it i.e.}, if $r$ is a multiple of $N$.  Then the unique solution to \suAC, up to conjugation, can be written as

\eqn\eeqoneg{ \widetilde g = w \; C_{\ell,N}, \qquad \widetilde h = z \;S_N\,,}
where $w$ and $z$ should be picked so these have determinant one, namely, they can each be any $N$th root of unity.  In principle we should sum over all such choices.  However, note there is a residual Weyl symmetry, namely, conjugation by $S_N$, which preserves the form of these matrices but has the effect of taking $w \rightarrow w \nu$.  Since $\nu$ generates all the $N$th roots of unity, by repeatedly applying this transformation we range over all these choices of $w$, so they are all Weyl-equivalent, and so we should only count one of them, say, $w=1$.\foot{The sum over different lifts $\tilde h$ of $h$ introduces a factor of $N$ which is canceled by a corresponding factor coming from the volume of the gauge group, as in \liftItilde.  Thus we may also specialize to a single choice of $z$, say, $z=1$.}

Now consider the second equation in \suAC.  Since $\nu^r = {\zeta_N}^{\ell r}=1$ and $w^r=1$, we find:

\eqn\ksols{\eqalign{
&\qquad\qquad \mu = {\tilde g}^r = {\epsilon_N}^{\ell r} \cr
&\qquad\qquad \;\;\;\;\Downarrow\cr
& k =\cases{
0 & $N \; {\rm odd}$ \cr
N/2 & $N \; {\rm even}$ , $r$ {\rm an odd multiple of }$N$ \cr
0 & $N \; {\rm even}$ , $r$ {\rm an even multiple of }$N$ }\,.
}
}
This is a generalization of the mod $4$ dependence of the solutions to~\suAC\
 we encountered in the $SU(2)$ case in section $3$.

Before proceeding to general $d$, let us see how to compute the contribution to the index from these holonomies.  Again consider an adjoint chiral multiplet.  The action of $\tilde g$ and $\tilde h$ is no longer diagonal in the usual weight basis, so we must pass to a different one.  Recall that if $e_i$ denotes the usual basis of $\C^N$ on which $SU(N)$ acts, then the usual weight basis for the adjoint is $E_{i,j}=e_i \otimes {e_j}^*$, where we should quotient out by the trace part $\sum_i E_{i,i}$.  However, the appropriate basis here will be

\eqn\Fbasis{
 F_{m,n} = \sum_j e^{2 \pi i m j/N} E_{j,j-n} \,.
}
where now we drop $F_{0,0}$, which represents the trace part.  Then we find

\eqn\nice{ {\widetilde g} \cdot F_{m,n} = e^{2 \pi i n \ell/N} F_{m,n}, \qquad\qquad
 {\widetilde h} \cdot F_{m,n} = e^{2 \pi i m/N} F_{m,n}\,.}
Thus the contribution to the index of an adjoint chiral multiplet is (noting the eigenvalue
 of ${ g}$ can be written as $e^{2 \pi i n \ell (r/N) /r}$, where $r/N$ is an integer by assumption),

\eqn\eigenva{ {\prod_{m,n}}' {\cal I}_\chi^{(R)}(\frac{n \ell r}{N},e^{2 \pi i m/N})\,,}
where the prime denotes that we exclude $m=n=0$ in the product.  As in the $SU(2)$ case, this contribution is an isolated point, and there is no further integral or sum that must be performed, although now there are $N-1$ inequivalent sectors labeled by $\ell$.

Now we return to the problem of finding solutions to \suAC\ for general $d=(\ell,N) $.  Note that for a given $d>0$ to contribute, $r$ must be a multiple of $N/d$ in order that $\nu^r=1$.  From \UNACsol\ we can write:

\eqn\UNACsol{
\widetilde g =  g_d \otimes C_{\ell/d,N/d}, \qquad\qquad
 \widetilde h =  h_d \otimes S_{N/d}\,,
}
Here $g_d$ and $h_d$ are $U(d)$ matrices, and in order that $\tilde g$ and $\tilde h$ have determinant $1$, these must have determinant an $(N/d)$th root of unity.  By conjugation we may take both to be diagonal, and write:

\eqn\gdhddef{ g_d= {\rm diag}(w_1,...,w_d), \;\;\; h_d = {\rm diag}(z_1,...,z_d) }
Now let us impose the second condition in \suAC.  In order for ${\tilde g}^r$ to be a scalar matrix, both ${g_d}^r$ and ${C_{\ell/d,N/d}}^r$ must be scalar matrices.  The latter condition follows already from $\nu^r=1$, and the former means that we can write:

\eqn\wdef{ w_i = e^{\frac{2 \pi i }{r} (m_i + \frac{\tilde k}{N})} }
where the $m_i$ are integers, and $\tilde k$ will be determined in a moment.  The condition that $g_d$ has determinant an $(N/d)$th root of unity gives:

\eqn\gdsum{ \sum_{i=1}^d m_i + \frac{ {\tilde k} d}{N} = 0 \; ({\rm mod} \; \frac{r d}{N}) }
For this to have solutions $\tilde k$ must be an integer multiple of $N/d$.  Then $\mu$ is given by:

\eqn\muac{ \mu = {\tilde g}^r ={\epsilon_{N/d}}^{\ell r/d} e^{\frac{2 \pi i \tilde k}{N}} }
which, after a short computation, means that we can write $k$ as

\eqn\ksolsgene{
\eqalign{
k =\cases{
\tilde k,& $N/d$ \ odd \cr
\tilde k +\frac{r d}{s} (s/2),  & $N/d$\ even
}\,.}}
From this formula we can see that, in general, there is a shift similar to the one we saw above.  This shift is by $s/2$ and occurs when $N/d$ is even and $\frac{r d}{s}$ is odd (recall $k$ is defined modulo $s$).  

Using \ksolsgene, we may pass between $\tilde k$ and $k$.  Note that this map depends on $\ell$ through $d=(\ell,N)$.  Then we can state the allowed solutions in terms of $\tilde k$ and $\ell$ very simply: since $\tilde k$ ranges over the multiples of $N/d$, we see the allowed pairs $(\tilde k,\ell)$ are precisely those for which

\eqn\allowed{ \tilde k\; \ell = 0 \; ({\rm mod} \; N)\,.}
Note this equation is consistent with the identification $\tilde k \sim \tilde k + s$ since $\ell$ is a multiple of $N/s$.

We should also be careful to fix the residual gauge symmetry that remains once we choose these forms for $\tilde g$ and $\tilde h$.  First, there are the Weyl symmetries which permute the elements of $g_d$ and $h_d$, which we can use to fix $m_1 \geq ... \geq m_N$.  In addition, there are Weyl symmetries which take the $i$th entry of the $n$th $d \times d$ block to the $i$th entry of the $(n+1)$th block, for any chosen $i$.  This has the effect of taking:

\eqn\blockweyl{ e^{\frac{2 \pi i}{r} (m_i + \frac{\tilde k}{N})} \rightarrow \nu e^{\frac{2 \pi i}{r}(m_i + \frac{\tilde k}{N})}  \Rightarrow m_i \rightarrow m_i + \frac{\ell r}{N} }
We can use this freedom to shift all of the $m_i$ to lie between $0$ and $\frac{r d}{N}-1$.  To summarize, the Weyl-inequivalent solutions to \suAC\ for a given $\mu={\zeta_N}^k$ and $\nu={\zeta_N}^\ell$ are given by:
\eqn\suacsummary{ \tilde g= {\rm diag}(e^{\frac{2 \pi i}{r}(m_1+\frac{\tilde k}{N})},...,e^{\frac{2 \pi i}{r}(m_d+\frac{\tilde k}{N})}) \otimes C_{\ell/d,N/d}, \;\;\; h_d = {\rm diag}(z_1,...,z_d) \otimes S_{N/d} }
where we sum over integers $m_i$ with $\frac{rd}{N} > m_1 \geq ... \geq m_d \geq 0$, satisfying \gdsum, with $\tilde k$ related to $k$ through \ksolsgene.

Now let us describe how to compute the contributions  $Z_{\mu,\nu}$
to the lens space index.  As before, to diagonalize the matrices $\tilde g$ and $\tilde h$ in the adjoint representation we must pass to an appropriate basis.  Such a basis turns out to be a tensor product of the usual weight basis and the basis used in \Fbasis, and is spanned by $F_{m,n;i,j}$, $m,n=1,..,N/d$, $i,j=1,...,d$.   We find
\eqn\appbasis{\widetilde g \cdot F_{m,n;i,j} = e^{2 \pi i n \ell/N + (m_i - m_j)/r} F_{m,n;i,j}, \qquad
 \widetilde h \cdot F_{m,n;i,j} = e^{2 \pi i m d/N} z_i {z_j}^{-1} F_{m,n;i,j} \,.}
Here the trace part is given by $\sum_i F_{0,0;i,i}$, and we should drop the corresponding eigenvalue of $1$.  Thus the contribution of an adjoint field from this holonomy is
\eqn\contadjsu{ \frac{\prod_{m,n=1}^{N/d} \prod_{i,j=1}^d {\cal I}_\chi^{(R)}([\frac{n \ell r}{N}+m_i - m_j],e^{2 \pi i md/N}z_i {z_j}^{-1})}{{\cal I}_\chi^{(R)}(0,1)} \,,}
where the denominator removes the contribution of the trace part.
The Haar measure is given here by

\eqn\suhaar{
\Delta_{{\bf m}}^\ell(z) =\frac{1}{|W_{{\bf m},N}|}\; \prod_{m,n=1}^{N/d} \prod_{i<j}^d ((1- e^{2 \pi i md/N} z_i {z_j}^{-1})(1- e^{2 \pi i md/N} {z_i}^{-1} z_j))^{\delta_{m_i - m_j+n \ell r/N,0}} \,.
}

Finally let us discuss how these sectors are arranged to form the partition functions of the various choices for the global group structure.  Recall in \readinglines\ it was argued there is a theory $(SU(N)/\Z_d)_n$ for $d$ a divisor of $N$ and $n=0,...,d-1$.\foot{For the remainder of this section we will reserve $d$ for the order of the subgroup $\Z_d$ by which we quotient $SU(N)$.}  These differ by adding a term of the following form to the action:

\eqn\heypont{ \frac{2 \pi i n}{d} \frac{P_2(w_2(k,\ell))}{2}\,,}
where $P_2/2$ is a certain squaring operation which takes the class $w_2(k,\ell) \in H^2(L(r,1) \times \S^1,\Z_d)$ to an element in $H^4(L(r,1) \times \S^1,\Z_d) \cong \Z_d$.  For the case $d=N$, we find it natural to conjecture:

\eqn\ptwodef{
\frac{P_2(w_2(k,\ell))}{2} = \tilde{k}\; \ell \in \Z_N \,,}
where $\tilde{k}$ is as defined by \ksolsgene.  One reason for this conjecture is the fact that the instanton number $\ell_{inst}$ satisfies\  \readinglines \WittenNV:

\eqn\instrel{
\ell_{inst} =\frac{1}{N} \frac{P_2(w_2(k,\ell))}{2}  \; ( {\rm mod} \; 1) \;.
}
Thus this conjecture is consistent with the observation that flat connections only exist when ${\tilde k}\; \ell = 0 \;({\rm mod} \; N)$.

For general $d$, note that only those bundles with $\tilde k$ and $\ell$ a multiple of $N/d$ lift to $SU(N)/\Z_d$ bundles, so that we may write:

\eqn\heyk{ \tilde k = \hat{k} \;\frac{N}{d}, \qquad\qquad\ell = \hat{\ell} \;\frac{N}{d} \,,}
where $\hat{k},\hat{\ell}$ live in $\Z_d$.  Then we find, using \ptwodef:

\eqn\ptwodeftwo{ \frac{P_2(w_2(\hat{k},\hat{\ell}))}{2} = \hat{k} \;\hat{\ell}\; \frac{N}{d} \in \Z_d \,.}
With this in mind, we can write the partition function for $(SU(N)/\Z_d)_n$ as:
\eqn\sunpartdef{
{\cal I}_{\left(SU(N)/\Z_d\right)_n} = \frac{1}{d} \sum_{\hat{k},\hat{\ell}=0}^{d-1} e^{2 \pi i n \hat{k} \hat{\ell}/d} Z_{\tilde{k}=\hat{k} N/d,\;\ell=\hat{\ell} N/d} \,.
}

Recall that the lens space index is not sensitive to the (ordinary) theta angle of the theory.  However, the choice of $(SU(N)/\Z_d)_n$ may be changed as we take $\theta \rightarrow \theta+2 \pi$, specifically, one finds, in the case $d=N$ \readinglines:

\eqn\Ttrans{
(SU(N)/\Z_N)_{n} \rightarrow (SU(N)/\Z_N)_{n+1} \;\;\;\; {\rm as} \;\; \theta \rightarrow \theta+2 \pi \,.
}
For general $d$, from \ptwodeftwo, we find that there is a shift of $n \rightarrow n+\frac{N}{d}$ as $\theta \rightarrow \theta+2 \pi$.  The lens space indices of theories related by this operation should be equal.  This is consistent with \sunpartdef\
since  the phase factor $e^{2 \pi i n \hat{k} \hat{\ell}/d}$ is invariant under $n \rightarrow n+N/d$ provided that $\ell \,\tilde{k} = 0\; ({\rm mod}\; N)$, which are the only values for which $Z_{\tilde{k},\,\ell}$ is non-zero.

\subsec{Sample computations for $su(N)$ ${\cal N}=4$ SYM}

Now let us turn to the theory we are interested with in this section, ${\cal N}=4$ SYM with gauge group based on Lie algebra $su(N)$.  As above, the field content consists of the ${\cal N}=1$ vector multiplet along with three ${\cal N}=1$ adjoint chiral multiplets, which can be weighed with fugacities $u_a$, $a=1,2,3$, satisfying $\prod_a u_a=1$.  Then the sectors $Z_{\mu,\nu}$ can be written most conveniently in terms of $\tilde{k}$ and $\ell$ as,

\eqn\zshsugen{
\eqalign{
&Z_{\tilde{k},\ell} = \sum_{m_i| \sum_i m_i = -\tilde{k} \,d/N ({\rm mod} \; r d/N)}\int_{\prod_i z_i = 1} \prod_i\frac{dz_i}{2 \pi i z_i}\Delta_{{\bf m}}^\ell(z)
\times \cr 
&\frac{\prod_{m,n=1}^{N/d} \prod_{i,j=1}^d {\cal I}_V([\frac{n \ell r}{N}+m_i - m_j],e^{2 \pi i m d/N}z_i {z_j}^{-1}) \prod_{a=1}^3 {\cal I}_\chi^{(\frac23)}([\frac{n \ell r}{N}+m_i - m_j],u_a
e^{2 \pi i m d/N}z_i {z_j}^{-1}) }{{\cal I}_V(0,1)\prod_{a=1}^3{\cal I}_\chi^{(\frac23)}(0,u_a)}\,.
}}

Let us now compute this in several examples.  For simplicity we specialize to $p=q=x$ and $u_a=1$.  As our first example, we compute the sectors $Z_{\tilde{k},\ell}$ for $su(3)$ and $r=3$ and arrange them in the table below, 

\eqn\SUthree{\eqalign{
&
\vbox{\offinterlineskip\tabskip=0pt
\halign{\strut\vrule#
&~$#$~\hfil\vrule
&~$#$~\hfil\vrule
&~$#$~\hfil\vrule
&~$#$~\hfil\vrule
&~$#$~\hfil\vrule
&~$#$\hfil
&\vrule#
\cr
\noalign{\hrule}
& \ell \backslash \tilde{k}  & 0 & 1 & 2   \cr
\noalign{\hrule}
& \;\;0   & a & b & b \cr
\noalign{\hrule}
& \;\;1   & c & 0 & 0 \cr
\noalign{\hrule}
& \;\;2   & c & 0 & 0 \cr
}\hrule}
}
}
where,

\eqn\suthreehols{ a = 4 + 6 x^{2/3} + 33 x^{4/3}+\dots,\;\;\; b =  3 + 9 x^{2/3} + 27 x^{4/3}+\dots, \;\;\; c=1 - 3 x^{2/3} + 6 x^{4/3} +\dots\,.}
The partition functions for the various choices of gauge groups are then computed from \sunpartdef\ as:

\eqn\SUthreeparts{\eqalign{
& {\cal I}_{SU(3)} = a = 4 + 6 x^{2/3} + 33 x^{4/3}+\cdots\,, \cr
& {\cal I}_{(SU(3)/\Z_3)_0} = Z_{(SU(3)/\Z_3)_1} = Z_{(SU(3)/\Z_3)_2} = \frac{1}{3} (a + 2b +2 c) = 4 + 6 x^{2/3} + 33 x^{4/3}+\cdots\,.
}}
These choices of gauge group are all related by various actions of the $S$-duality group, and their equality is a non-trivial test of this duality.

As an example with composite $N$, we compute the lens space index for $su(6)$ on $L(6,1) \times \S^1$.  The non-zero contributions can be arranged as

\eqn\SUsix{\eqalign{
&
\vbox{\offinterlineskip\tabskip=0pt
\halign{\strut\vrule#
&~$#$~\hfil\vrule
&~$#$~\hfil\vrule
&~$#$~\hfil\vrule
&~$#$~\hfil\vrule
&~$#$~\hfil\vrule
&~$#$~\hfil\vrule
&~$#$~\hfil\vrule
&~$#$\hfil
&\vrule#
\cr
\noalign{\hrule}
& \ell \backslash \tilde{k}  & 0 & 1 & 2  & 3 & 4 & 5\cr
\noalign{\hrule}
& \;\;0   & a & b & c & d & c & b \cr
\noalign{\hrule}
& \;\;1   & e & 0 & 0 & 0 & 0 & 0 \cr
\noalign{\hrule}
& \;\;2   & f & 0 & 0 & i & 0 & 0 \cr
\noalign{\hrule}
& \;\;3   & g  & 0 & h & 0 & h & 0 \cr
\noalign{\hrule}
& \;\;4   & f & 0 & 0 & i & 0 & 0 \cr
\noalign{\hrule}
& \;\;5   & e & 0 & 0 & 0 & 0 & 0 \cr
}\hrule}
}
}
where up to order $x^{\frac43}$ the different entries in the table have the 
following values

\eqn\susixresults{\eqalign{
a = 80 + 516 x^{2/3} + 2568 x^{4/3} +\dots ,\; &\;\;b = 75 + 531 x^{2/3} + 2520 x^{4/3} + \dots, \cr
c =78 + 522 x^{2/3} + 2556 x^{4/3}+\dots,\;\; &\;\;d= 76 + 528 x^{2/3} + 2526 x^{4/3}+\dots , \cr
e =1 - 3 x^{2/3} + 6 x^{4/3}+\dots,\;\;\;\;\;\;\;\;\;\;\;\;\; &\;\;f=2 - 6 x^{2/3} + 12 x^{4/3} +\dots ,\cr
g= 4 - 12 x^{2/3} + 42 x^{4/3}+\dots,\;\;\;\;\;\;\;\;\; &\;\;h = 3 - 9 x^{2/3} + 36 x^{4/3}+\dots, \cr 
i = 1 - 3 x^{2/3} + 6 x^{4/3}+\dots\;\;\;\;\;\;\;\;\;\;\;\;\;\;& 
}}

The partition functions  thus are (up to order $x^{\frac43}$)

\eqn\SUsixparts{\eqalign{
& {\cal I}_{SU(6)} = a = 80 + 516 x^{2/3} + 2568 x^{4/3}+\dots\,,\cr
& {\cal I}_{(SU(6)/\Z_2)_0}=Z_{(SU(6)/\Z_2)_1} = 
\frac{1}{2}(a+d+g) =80 + 516 x^{2/3} + 2568 x^{4/3}+\dots\,,\cr
& {\cal I}_{(SU(6)/\Z_3)_0}=Z_{(SU(6)/\Z_3)_1}= Z_{(SU(6)/\Z_3)_2} = \frac{1}{3}(a+2c+2f) =80 + 516 x^{2/3} + 2568 x^{4/3}+\dots\,,\cr
& {\cal I}_{(SU(6)/\Z_6)_0}=\dots= Z_{(SU(6)/\Z_6)_5} = \frac{1}{6}(a+2b+2c+d+2e+2f+g+2h+2i) = \cr
& \;\;\;\;\;\;\;\;\;\;\;\;\;\;\;\;\;\;\;\;\;\;\;\;\;\;\;\;\;\;\;\;\;\;\;\;\;\;\;\;\;\;\;\;\;\;\;\;\;\;\;\;= 80 + 516 x^{2/3} + 2568 x^{4/3}+\dots\,.
}}
Once again theses are all related under actions of the $S$-duality group, and as a result they are all equal.

Finally let us do an example where there are multiple $S$-duality orbits, namely, $su(4)$.  Since this is the same as $so(6)$, it also serves as a test of the $S$-duality for an $SO(N)$ gauge group.  We find

\eqn\SUfour{\eqalign{
&
\vbox{\offinterlineskip\tabskip=0pt
\halign{\strut\vrule#
&~$#$~\hfil\vrule
&~$#$~\hfil\vrule
&~$#$~\hfil\vrule
&~$#$~\hfil\vrule
&~$#$~\hfil\vrule
&~$#$~\hfil\vrule
&~$#$~\hfil\vrule
&~$#$\hfil
&\vrule#
\cr
\noalign{\hrule}
& \ell \backslash \tilde{k}  & 0 & 1 & 2  & 3 \cr
\noalign{\hrule}
& \;\;0   & a & b & c & b \cr
\noalign{\hrule}
& \;\;1   & d & 0 & 0 & 0  \cr
\noalign{\hrule}
& \;\;2   & e& 0 & f & 0 \cr
\noalign{\hrule}
& \;\;3   & d & 0 &  0 & 0 \cr
}\hrule}
}
}
where now up to order $x^{\frac43}$

\eqn\sufourresults{\eqalign{
 a = 10 + 30 x^{2/3} + 138 x^{4/3}+\dots, &\;\; b = 8 + 36 x^{2/3} + 120 x^{4/3}+\dots, \cr
c=9 + 33 x^{2/3} + 132 x^{4/3}+\dots, \;\;& \;\; d=1 - 3 x^{2/3} + 6 x^{4/3}, \cr
e =2 - 6 x^{2/3} + 18 x^{4/3}+\dots, \;\;\;\;\;\;& \;\; f=1 - 3 x^{2/3} + 12 x^{4/3}+\dots
}}
The partition functions are given by:
\eqn\SUfourparts{\eqalign{
& {\cal I}_{SU(4)} =a = 10 + 30 x^{2/3} + 138 x^{4/3}+\dots\,,\cr
& {\cal I}_{(SU(4)/\Z_2)_0}= \frac{1}{2}(a+c+e+f) = 11 + 27 x^{2/3} + 150 x^{4/3}+\dots\,,\cr
& {\cal I}_{(SU(4)/\Z_2)_1}=\frac{1}{2} (a+c+e-f) =10 + 30 x^{2/3} + 138 x^{4/3}+\dots\,,\cr
& {\cal I}_{(SU(4)/\Z_4)_0}=...=Z_{(SU(4)/\Z_4)_3} = \frac{1}{4} (a+2b+c+2d+e+f) = 10 + 30 x^{2/3} + 138 x^{4/3}\dots\,.\cr
}}
Here the $S$-duality group has two orbits~\readinglines, one containing $(SU(4)/\Z_2)_0$ and another containing all the other choices.  Indeed, we can see that the indices within a given orbit agree, although they disagree for the different orbits.

\

\

\newsec{Brief summary of the results and comments}

Let us summarize the results of this paper. We have discussed in some detail the 
computation of the lens space index for gauge theories with non-simply connected groups.
Important new ingredients of the computation, which do not appear in the case of the simply connected gauge groups,  are  the ``almost'' commuting holonomies. We have suggested how to define the computation for such holonomies by extending the known prescription for commuting ones.
The technology of computing the lens space index was then put to work in the cases of ${\cal N}=1$ 
gauge theories related by IR  Seiberg dualities and ${\cal N}=4$ conformal S-dualities. We have 
further discussed how lens space index distinguishes versions of  gauge theories differing by discrete
(topological) theta angle parameters. In particular in all cases considered we found perfect agreement with the results of~\readinglines: the lens indices satisfy intricate identities following from different duality
patterns depending on the discrete theta parameters and/or global properties of the gauge group.
The existence of such identities is a highly non-trivial test of the suggested dualities.

Let us mention several possible directions for future work. First, a better understanding of the
connection between the discrete theta parameters in presence of torsion cycles and their 
precise relation to the discrete electric and magnetic charges as discussed in~\refs{\WittenNV,\VafaTF} 
is desirable. It is likely that the presence of such torsion cycles may make the analysis more subtle than in the non-torsion case~\refs{\FreedYA,\WittenWY}. We believe such an understanding will clarify the extra identities that we found were satisfied by the index, {\it i.e.}, they may correspond to identities following from dualities of the indices refined with such discrete electric and magnetic charges. Alternatively, it should be interesting to consider such relation for the partition
functions on manifold with no torsion cycles and (at least) two non-torsion cycles, {\it e.g.} $T^2\times \S^2$, once the technology of computing such partition functions is developed.
It would be also interesting to study whether the lens space index can teach us something interesting 
about global properties of the flavor structure of  ${\cal N}=2$ theories related by
Gaiotto's dualities. Some aspects of the lens space index of these theories were recently studied in~\refs{\AldayRS,\RazamatJXA}.

On more technical note several gaps in computation of lens indices should be addressed. First, the
 prescription for computing lens indices should be extended to cases where
a holonomy for a global symmetry with global anomaly is turned on. This will provide  more  
refined checks of dualities than discussed here. It will be conceptually beneficial 
to understand this issue and the computation of the index in presence of almost commuting holonomies by explicitly localizing the path integral. Yet another issue in need of a better
understanding is the relation of the lens space index in the large $r$ limit and the $3d$ supersymmetric 
index of the dimensionally reduced theory.

\vskip 1cm

\noindent {\bf Acknowledgments:}

We would like to thank G.~Festuccia, A.~Kapustin, G.~Moore, Y.~Tachikawa, and M.~Yamazaki for useful discussions.  We are especially grateful to  O.~Aharony and N.~Seiberg for sharing with us their results and for numerous discussions, 
and to E.~Witten for important comments and discussions.
 SSR gratefully acknowledges support from the Martin~A.~Chooljian and Helen Chooljian membership
 at the Institute for Advanced Study. The research of SSR was also partially supported by
NSF grant number PHY-0969448.  The research of BW was supported in part by DOE Grant DE-SC0009988.

\

\appendix{A}{A computation of Stiefel-Whitney classes}

In this appendix we explicitly compute the Stiefel-Whitney classes for $SO(N)$ bundles over $\R \P^3 \times \S^1$, to verify the formulas in section 2, 3, and 4, which applied to general $L(r,1)$, in the special case $r=2$.\foot{We are grateful to E. Witten for suggesting the main arguments of this appendix.}

First let us review the cohomology of this space.  The cohomology ring of $\R \P^n$ with $\Z_2$ coefficients is generated by a $1$-cycle ${\bf x}$ subject to the relation ${\bf x}^{n+1}=0$.  Using the fact that $\S^1 \cong \R \P^1$, we can describe the cohomology ring of the product space $\R \P^3 \times \S^1$ as,

\eqn\cohomrpt{ H^2(\R \P^3 \times S^1, \Z_2) = \Z_2[{\bf x},{\bf y}]/ <{\bf  x}^4= {\bf y}^2 = 0 > \,,}
where ${\bf x}$ is the $1$-cycle in $\R \P^3$, which we have called $\gamma$ in section 2, and ${\bf y}$ is the $1$-cycle of $\S^1$.  In particular, the second cohomology group is $\Z_2 \oplus \Z_2$, generated by ${\bf x}^2$ and ${\bf x y}$.

Next we use the following fact: the total Stiefel-Whitney class of a direct sum of vector bundles over a manifold ${\cal M}$ is the product of the total Stiefel-Whitney class of each bundle.  In particular, if we can show a vector bundle is equivalent to a direct sum of line bundles, one simply needs to compute the product of the Stiefel Whitney classes of these line bundles, which have the form $1+w_1$, where $w_1$ is the first Stiefel-Whitney class.  To compute $w_1$ for a line bundle, suppose we give it a connection, and without loss we may take it to be an $O(1)$ connection.  Then for each cycle generating a $\Z_2$ factor in $H^1({\cal M},\Z_2)$, one can pick up a factor of $\pm 1$ in the line bundle as one traverses the cycle.  Then $w_1$ of the bundle is given by the sum of the generators corresponding to those cycles where this factor is $-1$.

Now consider such a line bundle with a connection on ${\cal M}=\R \P^3 \times \S^1$.  Let us write the holonomies of the line bundle around the cycles generated by ${\bf x}$ and ${\bf y}$ as $g$ and $h$, respectively, taking values in $\{\pm 1 \}$.  If we write $g=(-1)^{\epsilon}$ and $h=(-1)^{\delta}$, with $\epsilon,\delta \in \{0,1\}$, then the total Stiefel-Whitney class is given by,

\eqn\totswsline{ 1 + \epsilon\,{\bf  x} + \delta\,{\bf  y}\,.}
Now suppose we have a direct sum of $N$ line bundles, with a connection which is diagonal,{\it i.e.}, $g = ((-1)^{\epsilon_1},...,(-1)^{\epsilon_N})$, $h=((-1)^{\delta_1},...,(-1)^{\delta_N})$ with $\epsilon_i,\delta_i \in \{ 0, 1 \}$.  Then the Stiefel Whitney class of this bundle is,

\eqn\totsws{w = \prod_{i=1}^{N} (1 + \epsilon_i\,{\bf x} + \delta_i\,{\bf  y})\,.}

Let us now relate this to the solutions we found above, classified by $\mu$ and $\nu$.  Recall that $g^2=1$, so $g$ is gauge equivalent to a holonomy which is diagonal with $\pm 1$ entries.  For $h$ this is not the case. However, if we can change the connection so as to continuously deform $h$ while preserving $g\,h=h\,g$, it must lie on the same bundle.  In particular, if we can deform $h$ to a diagonal matrix with $\pm 1$ entries, then we can argue the bundle is equivalent to a direct sum of line bundles, and use the formula above.

For $\nu=1$, $g$ and $h$ could both be taken in the maximal torus, so we can continuously deform $h$ to the identity while preserving $g\,h=h\,g$.  Suppose there are $2m$ eigenvalues of $-1$ in $g$ (note this must be even since $g$ has determinant $1$), which implies that $\mu = {\tilde g}^2 = (-1)^m$.  Then we find,

\eqn\nuone{ w = (1 + {\bf x})^{2 m} = 1 + m \;{\bf x}^2 \,.}
where we recall the coefficients in $w$ are defined modulo $2$. 
On the other hand, for $\nu=-1$, recall from \almost that the solutions could be written as,

\eqn\solmum{ g = {\rm diag}( ...,1,-1,-1) , \qquad\qquad h = {\rm diag}(...,-1,1,-1)\,.}
Here, again, we can continuously deform $h$ so that the first $N-3$ entries are all $1$, and let us suppose again there are $2m$ eigenvalues of $-1$ in $g$ (including the final two).  Then we find,

\eqn\numone{ w = (1+{\bf x})^{2m-2}(1+{\bf y})(1+{\bf x})(1+{\bf x}+{\bf y}) =
 1 + m\, {\bf x}^2 + {\bf x}\,{\bf y} + {\bf x}^2\, {\bf y} + (m+1)\, {\bf x}^3 \,{\bf y} \,.}
Once again, $\mu=(-1)^m$.  To summarize, the dictionary between $(\mu,\nu)$ and the Stiefel-Whitney class is,

\eqn\secswc{ w_2 = \frac{1-\mu}{2}\; {\bf x}^2 + \frac{1-\nu}{2} \;{\bf x}\,{\bf y}\,, }
which agrees with the general correspondence of \ghmunumap.

\listrefs
\end